\documentclass[twocolumn,superscriptaddress]{revtex4-1} 

\usepackage{amsmath}
\usepackage{amssymb}
\usepackage{times}
\usepackage{graphicx}
\usepackage{epstopdf}
\usepackage{floatrow}
\usepackage{float}
\usepackage[caption=false]{subfig}
\usepackage{array}
\usepackage{multirow}
\usepackage{algorithm}
\usepackage{algpseudocode}
\usepackage{hyperref}
\usepackage{afterpage}
\usepackage[titletoc,title]{appendix}
\usepackage{ragged2e}
\begin{document}
\sloppy

\title{Effective optimization using sample persistence: A case study on quantum annealers and various Monte Carlo optimization methods}
\author{Hamed Karimi}
\affiliation{\emph{1QB Information Technologies (1QBit), 458-550 Burrard Street, Vancouver, BC \, V6C 2B5, Canada}}
\affiliation{\emph{Department of Computer Science, University of British Columbia, 2366 Main Mall, Vancouver, BC \, V6T 1Z4, Canada}}
\author{Gili Rosenberg}
\affiliation{\emph{1QB Information Technologies (1QBit), 458-550 Burrard Street, Vancouver, BC \, V6C 2B5, Canada}}
\author{Helmut G. Katzgraber}
\affiliation{\emph{1QB Information Technologies (1QBit), 458-550 Burrard Street, Vancouver, BC \, V6C 2B5, Canada}}
\affiliation{\emph{Department of Physics and Astronomy, Texas A\&M University, College Station, TX \, 77843-4242, USA}}
\affiliation{\emph{Santa Fe Institute, 1399 Hyde Park Road, Santa Fe, NM \, 87501, USA}}

\begin{abstract}
We present and apply a general-purpose, multi-start algorithm for improving the performance of low-energy samplers used for solving optimization problems. The algorithm iteratively fixes the value of a large portion of the variables to values that have a high probability of being optimal. The resulting problems are smaller and less connected, and samplers tend to give better low-energy samples for these problems. The algorithm is trivially parallelizable, since each start in the multi-start algorithm is independent, and could be applied to any heuristic solver that can be run multiple times to give a sample. We present results for several classes of hard problems  solved using simulated annealing, path-integral quantum Monte Carlo, parallel tempering with isoenergetic cluster moves, and a quantum annealer, and show that the success metrics and the scaling are improved substantially. When combined with this algorithm, the quantum annealer's scaling was substantially improved for native Chimera graph problems. In addition, with this algorithm the scaling of the time to solution of the quantum annealer is comparable to the Hamze--de Freitas--Selby algorithm on the weak-strong cluster problems introduced by Boixo et al. Parallel tempering with isoenergetic cluster moves was able to consistently solve 3D spin glass problems with 8000 variables when combined with our method, whereas without our method it could not solve any. 
\end{abstract}

\maketitle

\section{Introduction}
\label{sec:introduction}

When solving optimization problems, it is a common strategy to run a heuristic solver multiple times and keep the best solution found. If all of the found solutions are aggregated into a sample, one could ask if there is any additional information in this sample, aside from the solution with the best value. In this paper, we present results that suggest that it is indeed possible to use the sample more efficiently. In particular, the idea is that if we observe which variables have the same value in all solutions, and fix those variables to those values, we have fixed them to their values in at least one optimum with a high degree of confidence. Once this has been done, the remaining problem tends to be much smaller and simpler to solve.

Significant research has been done on solving Ising problems, and the equivalent quadratic unconstrained binary optimization (QUBO) problems,
\begin{align*}
\begin{aligned}
&\mbox{argmin}_s \left[ s^T J s + h^T s \right] \\
&\mbox{s.t.} \, s \in \{-1, 1\}^N \\
\end{aligned}
\quad \quad \mbox{OR} \quad \quad
\begin{aligned}
&\mbox{argmin}_x \left[ x^T Q x \right] \\
&\mbox{s.t.} \, x \in \{0, 1\}^N\,,
\end{aligned}
\end{align*}
using different solvers. Numerous well known \mbox{NP-hard} problems can be formulated in this way, such as the travelling salesman problem, the quadratic assignment problem, the maximum cut problem, the maximum clique problem, the set packing problem, and the graph colouring problem (for a selection of formulations, see \cite{lucas2014ising}).

An idea that is prevalent in genetic algorithms is that of iteratively looking at a pair of solutions to find the common part, which is then fixed, and searching the remaining solution space. Although it is more common to breed two solutions, some research has suggested that it can be beneficial to breed a larger pool of solutions \cite{eiben1994genetic, eiben1995orgy, lis1997multi, ting2005mean}.  In addition, some algorithms maintain a reference set of elite solutions (typically obtained by performing a local search), and rely on finding the variables that are often set to the same value in the elite solutions, the idea being that they are likely to be set to the same value in the optimum \cite{wang2011effective,wang2013backbone,zhang2004configuration}. Another related concept is that of short- and long-term memory, in which local search algorithms such as tabu 1-opt guide their future exploration based on information gained from past exploration \cite{glover1989tabu,glover1990tabu}. The closest existing work to the method described in this paper is Chardaire~et~al. \cite{chardaire1995thermostatistical}, which fixes variables whose values remain constant as the temperature is decreased during simulated annealing. 

Quantum annealers have recently become commercially available \cite{johnson2011quantum,bunyk2014architectural}. Manufactured by \mbox{D-Wave} Systems Inc., they are designed to heuristically find low-energy states of an Ising problem. It has been suggested that quantum annealers have an advantage over classical optimizers due to quantum tunnelling, which allows an optimizer to search the solution space of an optimization problem by passing through energy barriers instead of traversing them. For certain problem classes, this might provide a quantum speedup \cite{ray1989sherrington,finnila1994quantum,kadowaki1998quantum,santoro2002theory,battaglia2005optimization,heim2015quantum,muthukrishnan2015tunneling}. For this reason, there has been much recent interest in benchmarking quantum annealers against classical solvers, although conclusive evidence of quantum speedup has not been shown to date \cite{boixo2014evidence,denchev2015computational,mcgeoch2013experimental,katzgraber2014glassy,king2015performance,king2015benchmarking,mandra2016strengths,ronnow2014defining,hen2015probing,katzgraber2015seeking, king2017quantum, mandra2017pitfalls}. The classical solvers that are most usually benchmarked against quantum annealers are simulated annealing and simulated quantum annealing, both of which were included in our benchmarks. 

Our research is based on an idea first proposed specifically for use with quantum annealers \cite{karimi2017boosting}, and studied only on a narrow problem set. In this work, we utilize a modified and improved version of the algorithm outlined in \cite{karimi2017boosting}, and show that it is effective for a range of solvers and hard problem sets. The original algorithm has several limitations that are mitigated in the method presented in this work (see Section~\ref{sec:multi_start_SPVAR}). The main focus of our study is not the benchmarking of the solvers against one another, but to gain an understanding for which problems and samplers our method is effective. 

\section{Method}
\label{sec:method}

In this paper, we study a multi-start version of the sample persistence variable reduction algorithm (SPVAR)  recently proposed by Karimi and Rosenberg \cite{karimi2017boosting}. For completeness, we briefly describe that algorithm in Section~\ref{sec:SPVAR}, after which, in Section~\ref{sec:multi_start_SPVAR}, we describe in more detail the multi-start variant that we used in this work. In Section~\ref{sec:special_considerations}, we discuss special considerations related to our method.

\subsection{SPVAR}
\label{sec:SPVAR}

\begin{figure*}[!htbp]  
   \centering
	\subfloat[]{
       \includegraphics[width=0.45\textwidth]{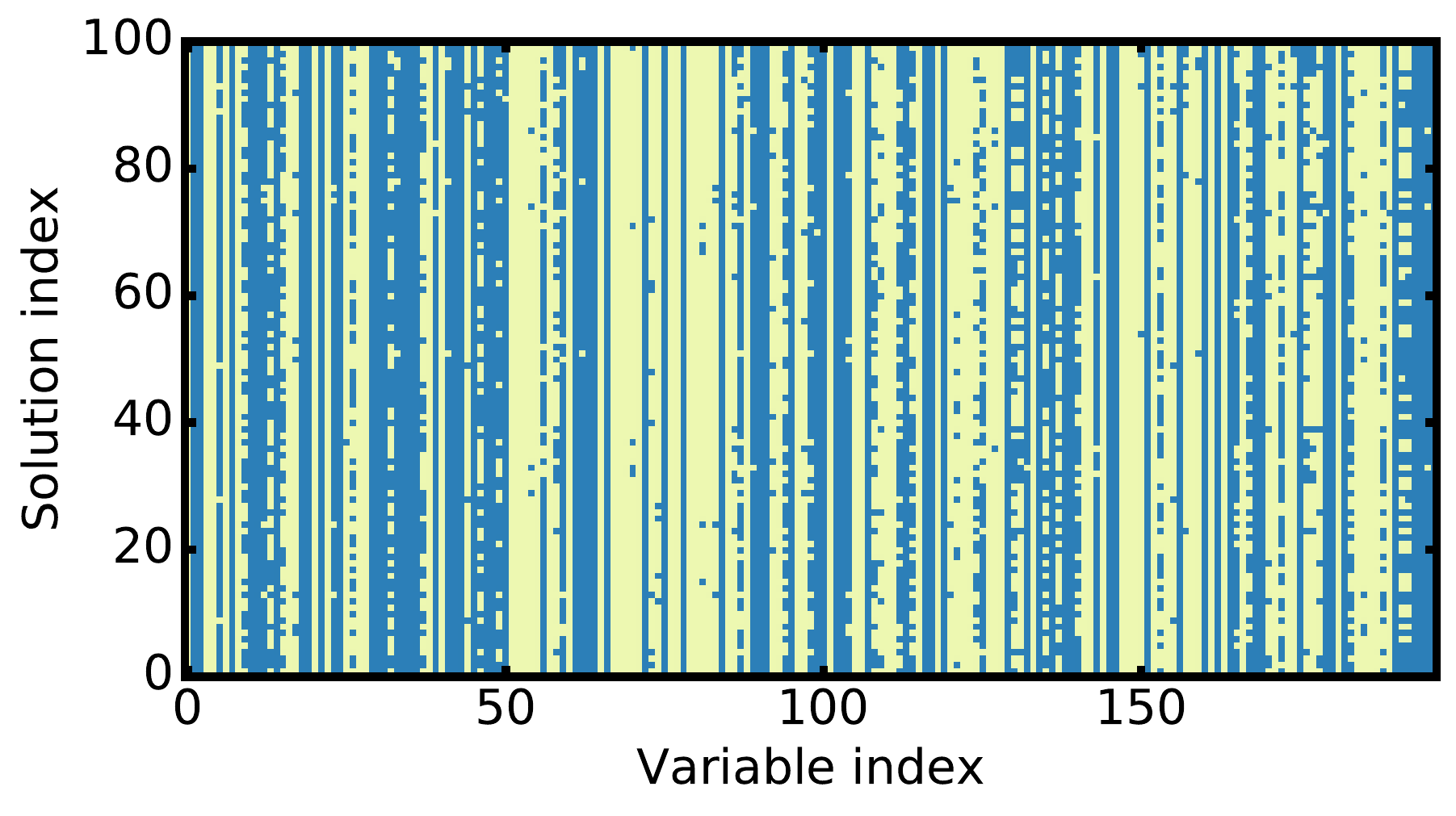}
       \label{fig:full_sample}}
   \quad
   \subfloat[]{
       \includegraphics[width=0.45\textwidth]{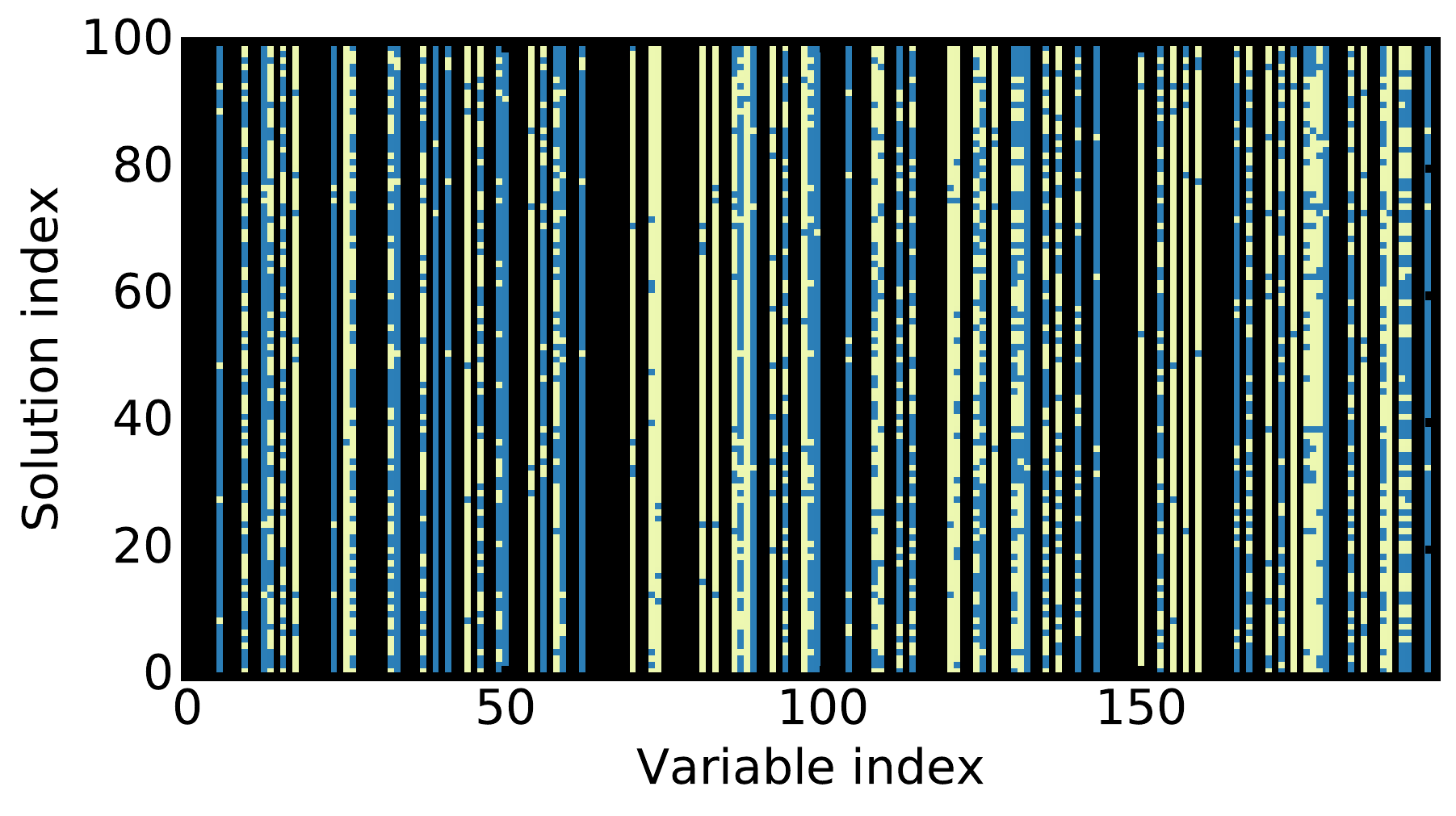}
       \label{fig:fixed_sample}
	}\\
   \subfloat[]{
   	\includegraphics[width=0.45\textwidth]{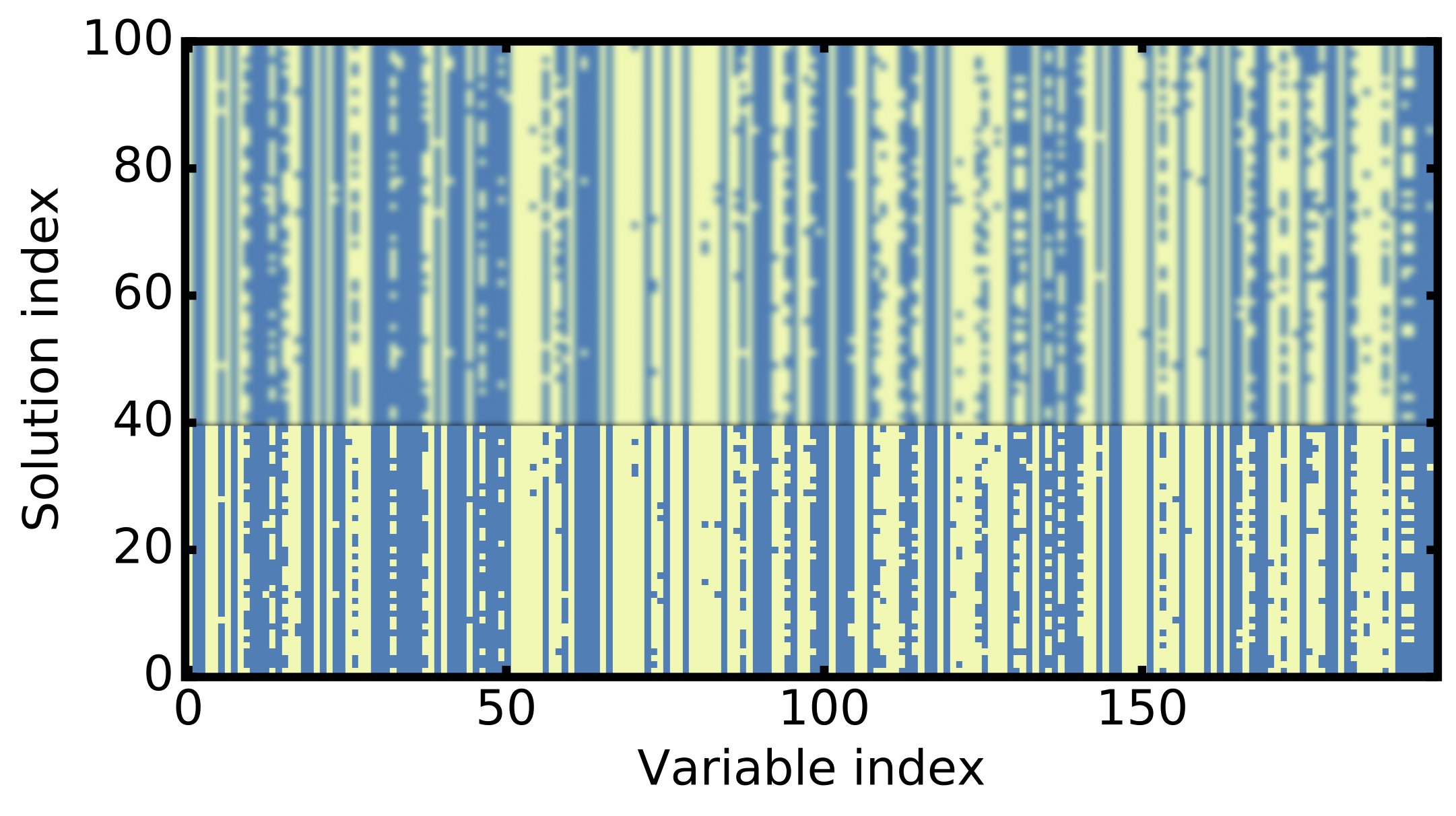}
     \label{fig:elite_sample}
	}
   	\quad
   \subfloat[]{
       \includegraphics[width=0.45\textwidth]{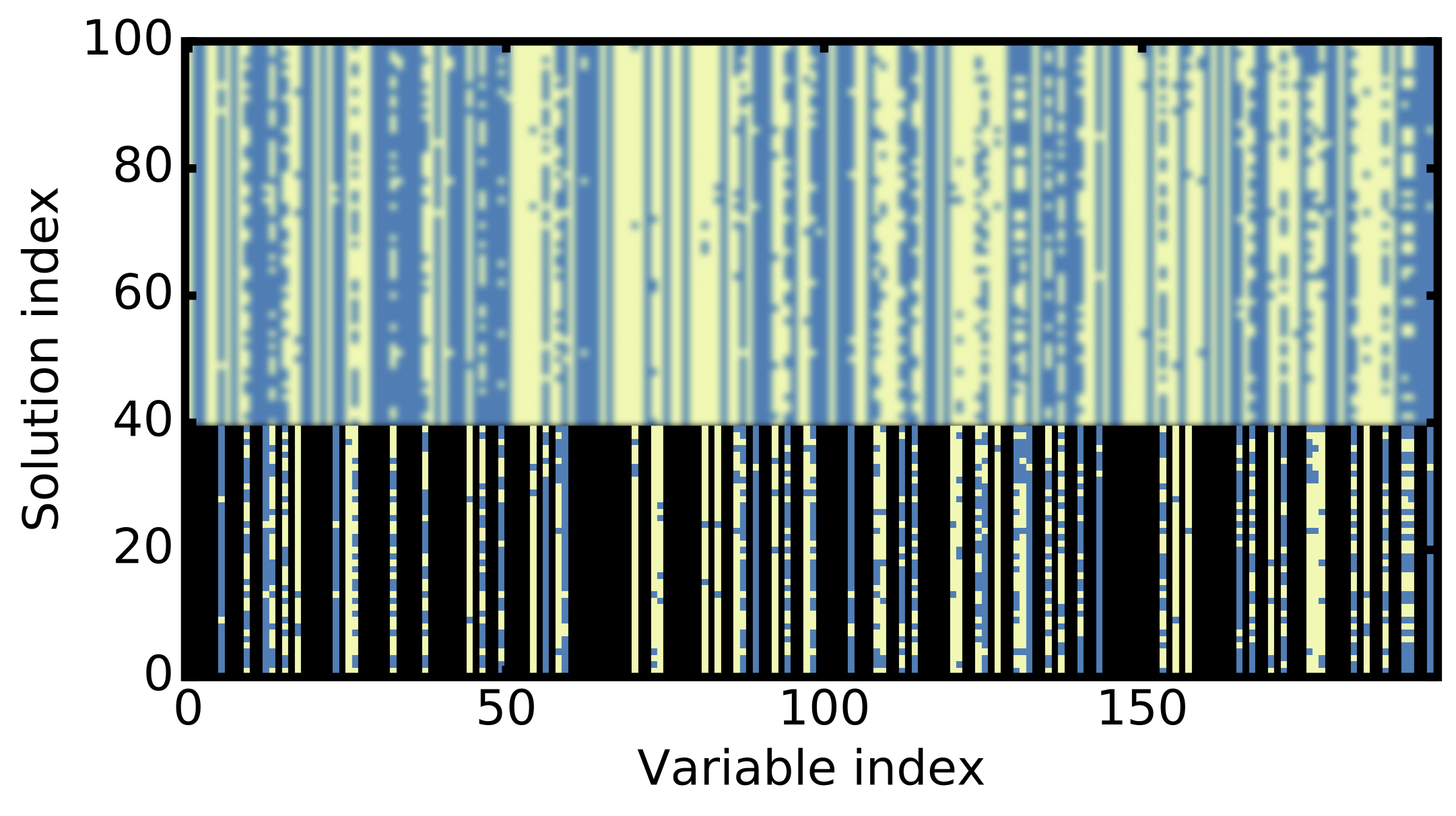}
     \label{fig:fixed_elite_sample}
	}
   \caption{A graphical illustration of the SPVAR algorithm. ({\bf a}) The full sample, with 100 solutions (the rows) and 200 variables (the columns). In each row, a dark dot (blue) indicates that that variable was $+1$ in that solution, and a light dot (beige) indicates that that variable was $-1$. ({\bf b}) The full sample, but with variables that had the same value in all solutions marked in black. ({\bf c}) The elite sample, formed by keeping only the 20\% lowest energy solutions. ({\bf d}) The elite sample, but with variables that had the same value in all solutions in the elite sample marked in black. In all figures, the solution index is sorted with respect to the energy, such that the solutions with the lowest energies are at the bottom. }
\label{fig:sample_barcode}
\end{figure*}

The SPVAR algorithm is based on the idea that, if a variable has the same value in all of the states obtained from a low-energy sampler, then it is more likely that the variable has that value in (at least) one optimum. This algorithm has two parameters: \textit{fixing\textunderscore threshold}, which controls what percentage of the solutions should share a value for a variable to be fixed; and \textit{elite\textunderscore threshold}, which controls the fraction of the low-energy part of the sample on which the algorithm should be applied. The idea is illustrated in Figure~\ref{fig:sample_barcode}. For further details, see Algorithm~\ref{algorithm:SPVAR} and the original paper \cite{karimi2017boosting}.

It is worth mentioning that finding and fixing these variables in a given sample is a very cheap computational operation, but as our results will show (Section~\ref{sec:results}), it significantly improves the performance of the underlying sampler. Using the SPVAR algorithm in a single run has some drawbacks, the most important being that, as this method is a heuristic method, there is a probability that some of the variables will be fixed to a value that does not occur in any optimal solution. This was the motivation for using this method in a multi-start fashion, as we describe in more detail in the following section.

\begin{algorithm}[!htbp]  
\footnotesize
\begin{algorithmic}
\Require Ising problem ($J,h$), \textit{sampler}, \textit{fixing\_sample\_size}, \textit{fixing\_threshold}, 
\State \textit{elite\_threshold} 
\\
\State Obtain sample of \textit{fixing\textunderscore sample\textunderscore size} from \textit{sampler}
\State Record energies from sample
\State Narrow down solutions to \textit{elite\textunderscore threshold} percentile
\State Find mean value of each variable in all solutions
\State Fix variables for which mean absolute value is larger than \textit{fixing\textunderscore threshold}
\State Update $J$ and $h$ \\
\Return $J$, $h$, recorded energies, and a mapping from fixed variables to values to which they were fixed 
\
\end{algorithmic}
\caption{\, SPVAR}
\label{algorithm:SPVAR}
\end{algorithm}
%

\subsection{Multi-start SPVAR}
\label{sec:multi_start_SPVAR}

The Multi-start SPVAR algorithm consists of \textit{num\textunderscore starts} independent starts of the SPVAR algorithm, using a sample size of \textit{fixing\textunderscore sample\textunderscore size}. Each of these starts is followed by calling the sampler on the modified problem, with a sample size of  \mbox{\textit{solving\textunderscore sample\textunderscore size}}. The output of the algorithm is a collection of all of the energies encountered during this process. The steps of the Multi-start SPVAR algorithm are summarized in Algorithm~\ref{algorithm:multi_start_SPVAR}. For considerations related to parameter choice, see \cite{karimi2017boosting}.  

Multi-start SPVAR has several advantages over single-start SPVAR. Firstly, running SPVAR has a finite but unknown probability of fixing variables incorrectly. This is mitigated by restarting the algorithm multiple times. Secondly, this version allows one to choose the fraction of the sample size that will be used for fixing variables (via SPVAR), out of the total sample size. This could be done with a single start as well, but doing so often results in a wastefully large sample being used for fixing variables. In addition, this algorithm is trivially parallelizable, allowing for a speedup if multiple samples can be collected in parallel. Finally, for problems with a degenerate ground state, SPVAR might fix variables to their values in an optimum, but contrary to their values in other optima. This would make it impossible to observe these other optima in the modified problem. Multiple restarts make the resulting sample less biased.

We remark that ISPVAR, the iterative version of SPVAR, could also be used as the building block for this  algorithm. The iterative algorithm can be useful if the objective is to fix more variables, and simplify the problem more drastically. However, every additional step of SPVAR that is applied incurs additional risk of fixing variables incorrectly. If a variable is fixed incorrectly at an early step, it is not unfixed at a later step. We suggest, instead, to set the thresholds more aggressively, which also results in the fixing of more variables. The price again is an increased risk of fixing variables incorrectly, but this risk can be mitigated by increasing \textit{num\textunderscore starts}. 

\begin{algorithm}[!htbp]  
\footnotesize
\begin{algorithmic}
\\
\Require Ising problem ($J, h$),  \textit{sampler}, \textit{fixing\_sample\_size}, 
\State \textit{solving\_sample\_size}, \textit{fixing\_threshold}, \textit{elite\_threshold}, \textit{num\_starts} \\
\State [Optional] Apply pre-processing to find modified \mbox{$J$, $h$}
\For{each start of \textit{num\textunderscore starts}}
		\State Apply SPVAR with sample of \mbox{\textit{fixing\textunderscore sample\textunderscore size}} to find modified $J$, $h$
		\State Record energies from sample
		\State [Optional] Apply pre-processing to find modified $J$, $h$
		\State [Optional] Fix variables via correlations to find modified $J$, $h$
	\State Obtain sample of size \textit{solving\textunderscore sample\textunderscore size} for modified $J$, $h$
	\State Record energies from sample
\EndFor \\
\Return Recorded energies, and a mapping from fixed variables to values to which they were fixed

\end{algorithmic}
\caption{\, Multi-start SPVAR}
\label{algorithm:multi_start_SPVAR}
\end{algorithm}

%
\subsection{Special considerations}
\label{sec:special_considerations}

For Ising problems with zero bias (such as those in \mbox{Section~\ref{sec:results_chimera}} and Section~\ref{sec:results_3Dspin}), there is a two-fold degeneracy of all states, which leads to many states appearing in a given sample with their reversed state. For this reason, if the method is applied exactly as described in Section~\ref{sec:multi_start_SPVAR}, it will generally fix no variables and fail to result in an improvement. As suggested in \cite{karimi2017boosting}, a possible solution is to break the degeneracy by arbitrarily fixing a single variable. However, as was found in that work, it is possible to use correlations in the sample to fix a cluster of correlated variables instead, which is the method we employed in this study. 

In cases in which the modified problem (after applying \mbox{SPVAR}) was disconnected, we took advantage of this fact to boost our results. In such cases, it is wasteful to assess the sampler's performance based on the energies for the whole problem, for the reason that even if the sampler did not manage to solve all of the disconnected problems at once, it may have solved all of them at least once. For this reason, we separated the solutions in a given sample into a partial sample for each connected problem. We then evaluated the partial energies for each connected component, for each solution; sorted the partial sample based on the partial energies; merged all of the partial (but sorted) solutions; and, finally, summed the partial energies to find the new sample. In this way, if the sampler solved each connected component in at least one solution, the best solution in the new sample would be a ground state of the whole problem. 

The optional pre-processing step can encompass many different methods for fixing variables. In this case, we used the \textit{fix\textunderscore variables} function in D-Wave's SAPI 2. We then added an additional method which deals efficiently with trees. For nodes (variables) with degree one, if the absolute value of the coupling is smaller than the absolute value of the bias for that variable, the value of the variable is determined by the sign of the bias. On the other hand, if the absolute value of the coupling is larger than the absolute value of the bias, the value of the variable is determined by the sign of the product of the coupling and the value of the neighbouring variable. This method can be applied recursively to fix, in the former case, or infer, in the latter case, the values of variables in trees, leading to a reduction in the number of variables equal to the number of variables in the tree. 

\section{Results and discussion}
\label{sec:results}

This section is organized as follows. In Section~\ref{sec:results_procedure}, we describe the benchmarking procedure we used in order to collect the results that are presented in this section. In the  sections that follow, we present results for different problem sets and solvers, and discuss them. In Section~\ref{sec:results_weak_strong}, we present results for weak-strong cluster problems. In Section~\ref{sec:results_chimera}, we present results for reduced-degeneracy Chimera graph problems with zero and non-zero bias; in Section~\ref{sec:results_3Dspin}, we present results for 3D spin glass problems; and in Section~\ref{sec:results_fault_diagnosis}, we present results for fault diagnosis problems. In Section~\ref{sec:results_maxksat}, we present results for Max-2-SAT and Max-3-SAT problems.

\subsection{Benchmarking procedure}
\label{sec:results_procedure}

For each problem set and each parameter value (when applicable), we generated or procured 50--100 instances, which were solved using a variety of samplers, with and without Multi-start SPVAR. We used the following heuristic solvers as samplers in this study: `SA'---an implementation of simulated annealing \cite{isakov2015optimised}; `SQA'---an implementation of discrete-time path-integral quantum Monte Carlo as a simulation of the quantum annealing process, which we refer to as SQA in this study \cite{kadowaki1998quantum,heim2015quantum,martovnak2002quantum}; `DW'---D-Wave's quantum annealer (we had access to the DW2X\textunderscore SYS4 and the DW\textunderscore 2000Q); and `PTICM'---parallel tempering Monte Carlo \cite{katzgraber2009introduction,geyer1992constrained,hukushima1996exchange} with isoenergetic cluster moves, also known as \emph{borealis} \cite{zhu2015efficient, zhu2016borealis}. In each case, the version that used Multi-start SPVAR is indicated by the suffix `SPVAR', for example, `SA\textunderscore SPVAR'. 

In addition, we solved, or procured best known solutions for, all problem instances with an additional solver which either guarantees finding the optimum, or finds it with a high probability. Specifically, the reduced-degeneracy Chimera graph problems were solved with the Hamze--de Freitas--Selby (HFS) algorithm \cite{selby2013github}, and the Max-$k$-SAT problems were solved with the CCLS \cite{luo2015ccls} and \emph{ahmaxsat} \cite{abrame2015ahmaxsat} algorithms. For the 3D spin glass problems and the fault diagnosis problems, we procured best known solutions using PTICM. For the 3D spin glass problems with planted solutions, the ground state was known \cite{wang2017spin}.

Using the energies returned by Multi-start SPVAR, we evaluated several success metrics: `Fraction of problems solved'; `Gap'---the difference in value between the best energy found and the best known solution's energy; `Residual'---the relative difference (in percent) in value between the best energy value found and the best known solution's energy; and `R99'---the size of sample required to observe the ground state with 99\% confidence. When calculating R99 values for results obtained with SPVAR, care is needed to get the correct value. We first calculated the mean success rate for each start and the number of starts required to observe the ground state with 99\% confidence. We then multiplied this value by the total size of the sample used in each start, to finally obtain the R99 value. 

\subsection{Weak-strong cluster problems}
\label{sec:results_weak_strong}

The weak-strong cluster problems were introduced by Boixo et al. \cite{boixo2016computational} as a toy model for studying the role of multiqubit tunnelling in a quantum annealer. They are Chimera graph problems in which the variables in each unit cell are ferromagnetically coupled to one another, and all have an equal bias which is $+1$ for the ``strong'' clusters and $h_w<0$ for the ``weak'' clusters. For $h_w=-0.5$, the ground state is doubly degenerate, corresponding to a state in which the weak cluster is either aligned with, or opposite to, the strong cluster. The case studied is  $-0.5<h_w<0$, which exhibits a false minimum in which the weak cluster is the opposite of the strong cluster. It has been shown, both theoretically and empirically, that SA tends to end up in the false minimum, unless cluster moves are allowed, whereas SQA and QA are able to tunnel out of the false minimum. 

The weak-strong problems were subsequently studied by Denchev et al. \cite{denchev2015computational}. It was shown that, on these problems, QA outperforms SA both in time and in scaling, and outperforms SQA in time (but not in scaling). Mandr\`a et al. \cite{mandra2016strengths} subsequently showed that these problems can, in fact, be solved more efficiently both in time and scaling by algorithms such as HFS and others. Later Mandr\`a et al. \cite{mandra2017pitfalls} also showed that these problems can be solved exactly in polynomial time, due to the planar structure of the logical problem. Our motivation for studying these problems was to check whether the application of our method would close the performance gap between the quantum annealer and HFS. 

We obtained the original problem instances from the above studies \cite{denchev2015computational, mandra2016strengths}. Since those instances were constructed for a different D-Wave chip, and each chip has a small number of arbitrarily located inactive qubits, we modified the instances to account for this. In addition, due to our having a limited amount of quantum annealer time, we had to change the weak bias $h_w$ from $-0.44$ (as in the original instances) to $-0.42$, which results in problems that are slightly easier for the quantum annealer. We expect that our results would hold qualitatively also for the original instances, run on the same chip they were benchmarked on in the past studies. 

\begin{figure}[!htbp]  
\captionsetup{width=.9\columnwidth}
   \centering
    \includegraphics[width=1.0\linewidth]{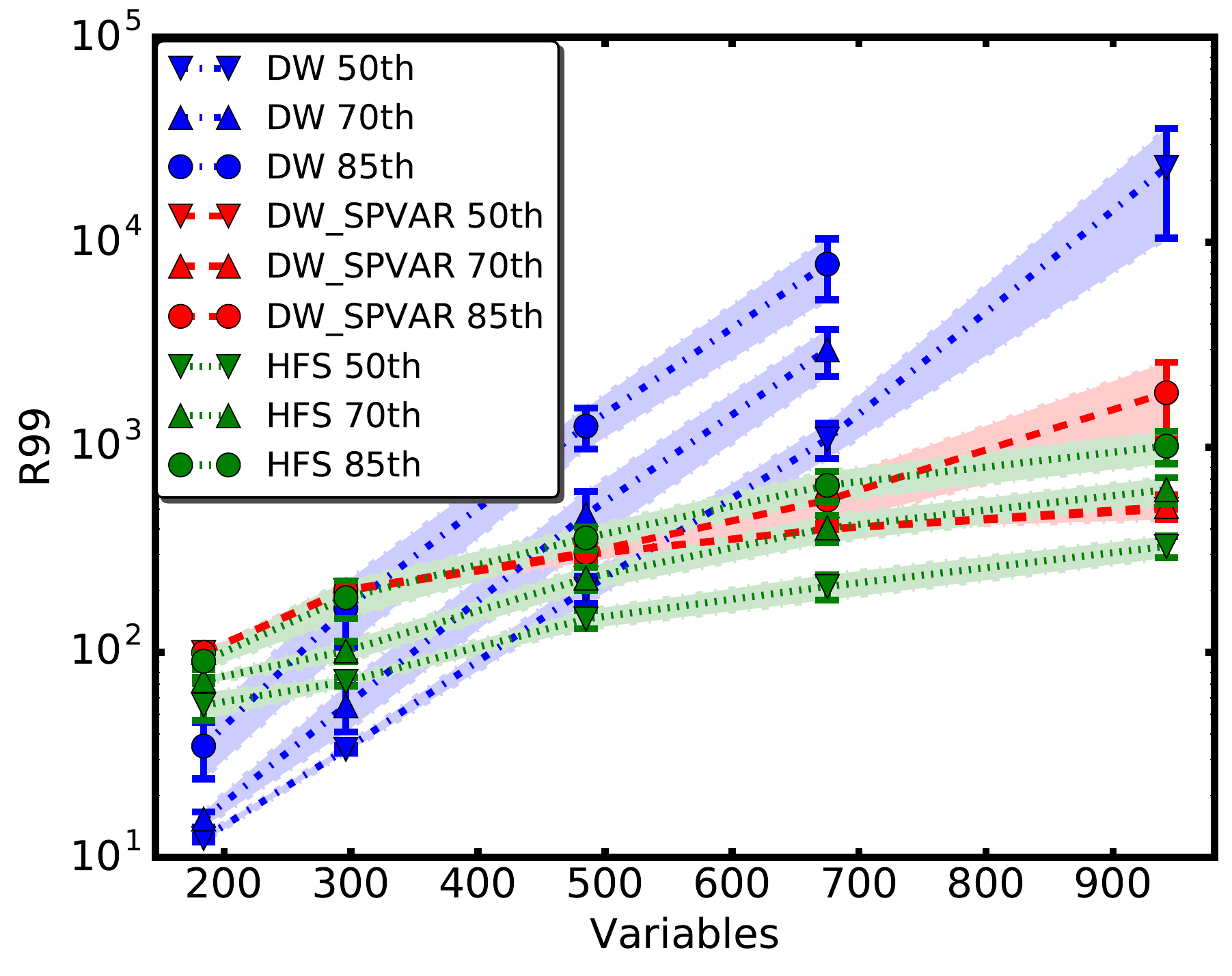} 
    \caption{Results for weak-strong cluster problems. We present the R99 values for different samplers, as a function of the problem size (number of variables), for different percentiles. The error bars were calculated 
     using bootstrapping. 
    }
    \label{fig:weak_strong_R99}
\end{figure}

We present the R99 values for HFS and for the quantum annealer used with and without our method in Figure~\ref{fig:weak_strong_R99}. In this case, the R99 value reported is the time required to find the ground state with 99\% confidence, in units of 20 $\mu$s, which was the annealing time we used for the quantum annealer. We were unable to solve these problems with SA or SQA due to our having insufficient resources.  The quantum annealer we used in this section was the DW2X\textunderscore SYS4 \footnote{\protect{The chip at our disposal has 1100 active qubits, a working temperature of 26 $\pm$5 mK, and a minimum annealing time of 20 $\mu$s.}}.

Our results  show a clear improvement in the R99 values when running the quantum annealer with our method, compared to running it without our method. Notably, the improvement is of two orders of magnitude for the largest problems. Furthermore, the scaling of the R99 value with the number of variables is clearly significantly improved, making it qualitatively comparable to HFS's scaling, and the R99 values themselves are also comparable with those of HFS.  We expect that qualitatively similar results would be drawn for the problems benchmarked in \cite{denchev2015computational, mandra2016strengths}. 

\subsection{Chimera graph problems}
\label{sec:results_chimera}

\subsubsection{Reduced-degeneracy Chimera graph problems}

Random Chimera graph problems have been benchmarked thoroughly in the past \cite{mcgeoch2013experimental,katzgraber2014glassy,king2015benchmarking,ronnow2014defining,
katzgraber2015seeking}, after which it was shown that they are not expected to show a quantum speedup \cite{katzgraber2014glassy}. However, they remain a well studied testbed, and are easy to construct. It was suggested that harder instances could be constructed by choosing the couplers and biases from a set with a large range compared to the spacing. Our instances were constructed by choosing the couplers and biases from a uniform probability distribution over the set $\{-n, \mbox{$-(n-1)$}, \mbox{$-(n-2)$}, \mbox{$n-2$},n-1,n \}$, which we denote by $U_{n,n-1,n-2}$. Such instances challenge a noisy quantum annealer, due to the effects of intrinsic control error (ICE). In addition, they challenge our method, since the mean change in energy due to an incorrectly flipped variable is much larger than in an instance in which the couplers and biases are chosen from the complete range $-n$ to $n$. 

In addition, in the interest of creating hard instances, it is helpful to reduce the ground state degeneracy \cite{katzgraber2015seeking}. To this end, we eliminated local degeneracies; doing so does not guarantee that the ground state will not be degenerate, but reduces the probability of this occurring. An instance was created via the following process. First, an initial set of couplers and biases was selected. Then, for each variable, we checked if there is a possible configuration of the neighbouring variables that would result in the effective field on the central variable being zero. If such a configuration was found, one of the couplers was changed to a different and randomly chosen value. This process was repeated until no more local degeneracies remained. We remark also that for small $n$, the probability of local degeneracies occurring is significant, but it falls rapidly as $n$ is increased. 

\begin{figure*}[!htbp]  
    \centering
    \subfloat[$h \neq 0$]{
        \includegraphics[width=0.45\textwidth]{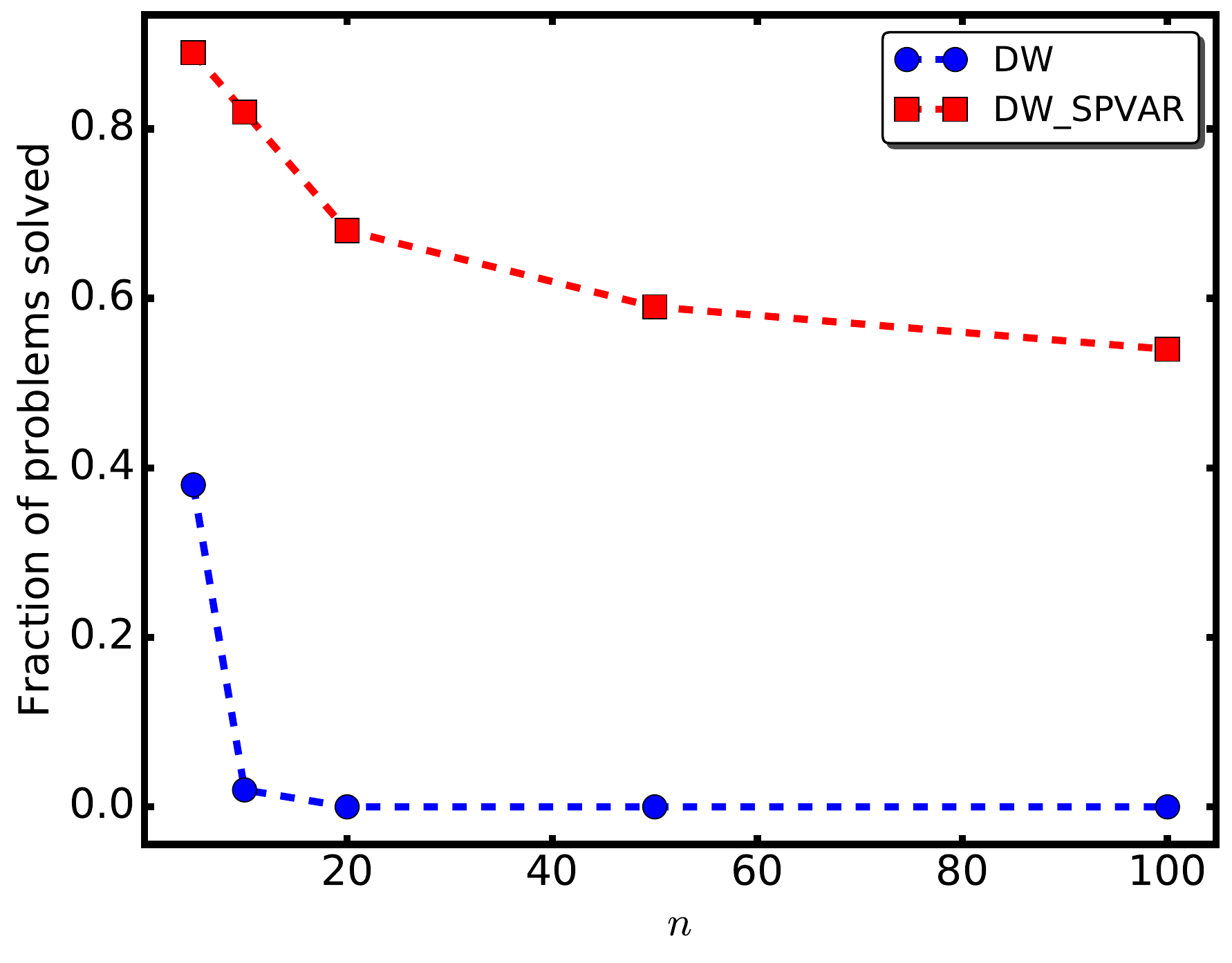} 
	\label{fig:chimera_fraction_problems_solved} 
	}
    \quad
    \subfloat[$h=0$]{
        \includegraphics[width=0.45\textwidth]{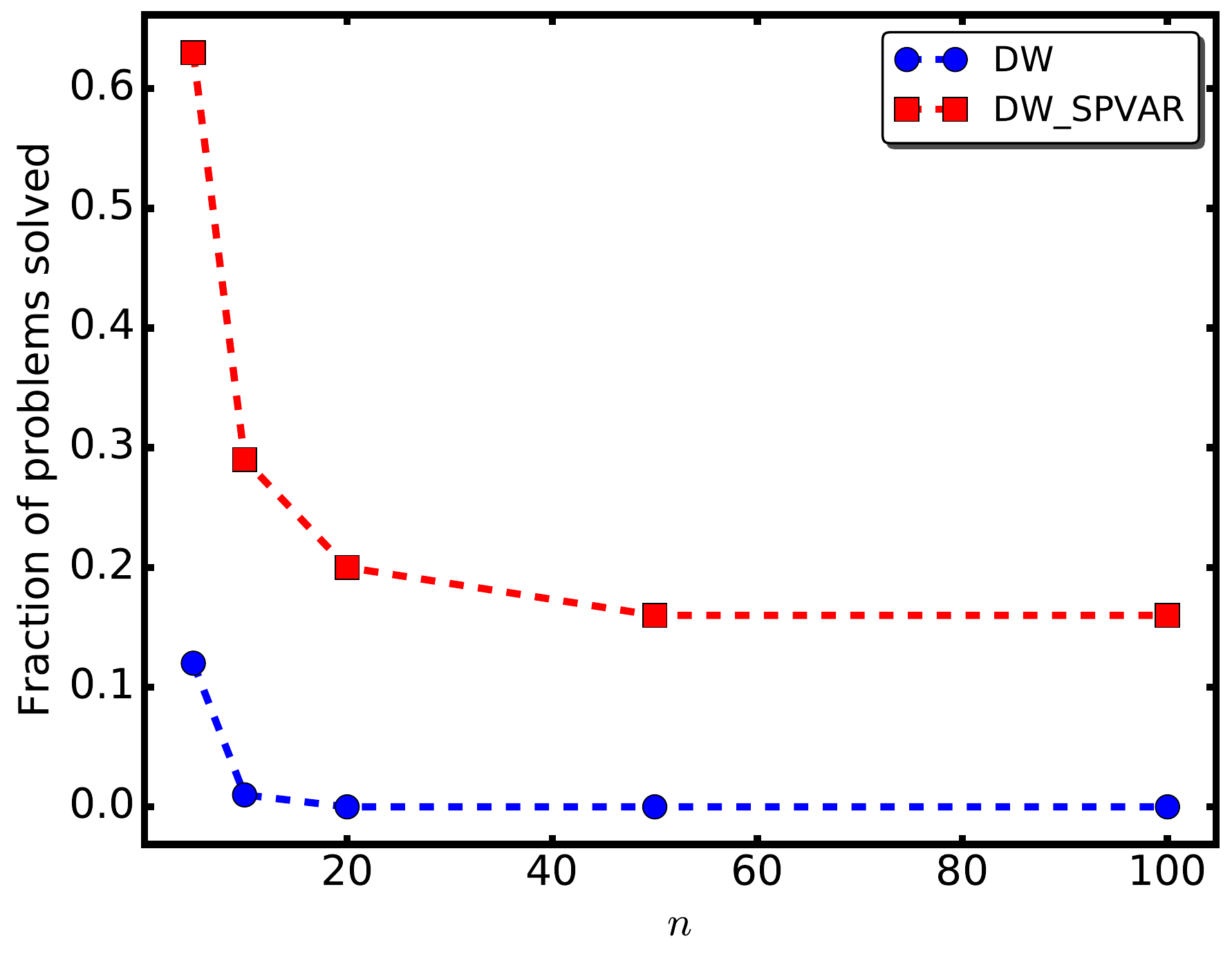} 
	\label{fig:chimera_h0_fraction_problems_solved} 
	} \\
        \subfloat[$h \neq 0$]{
         \includegraphics[width=0.45\textwidth]{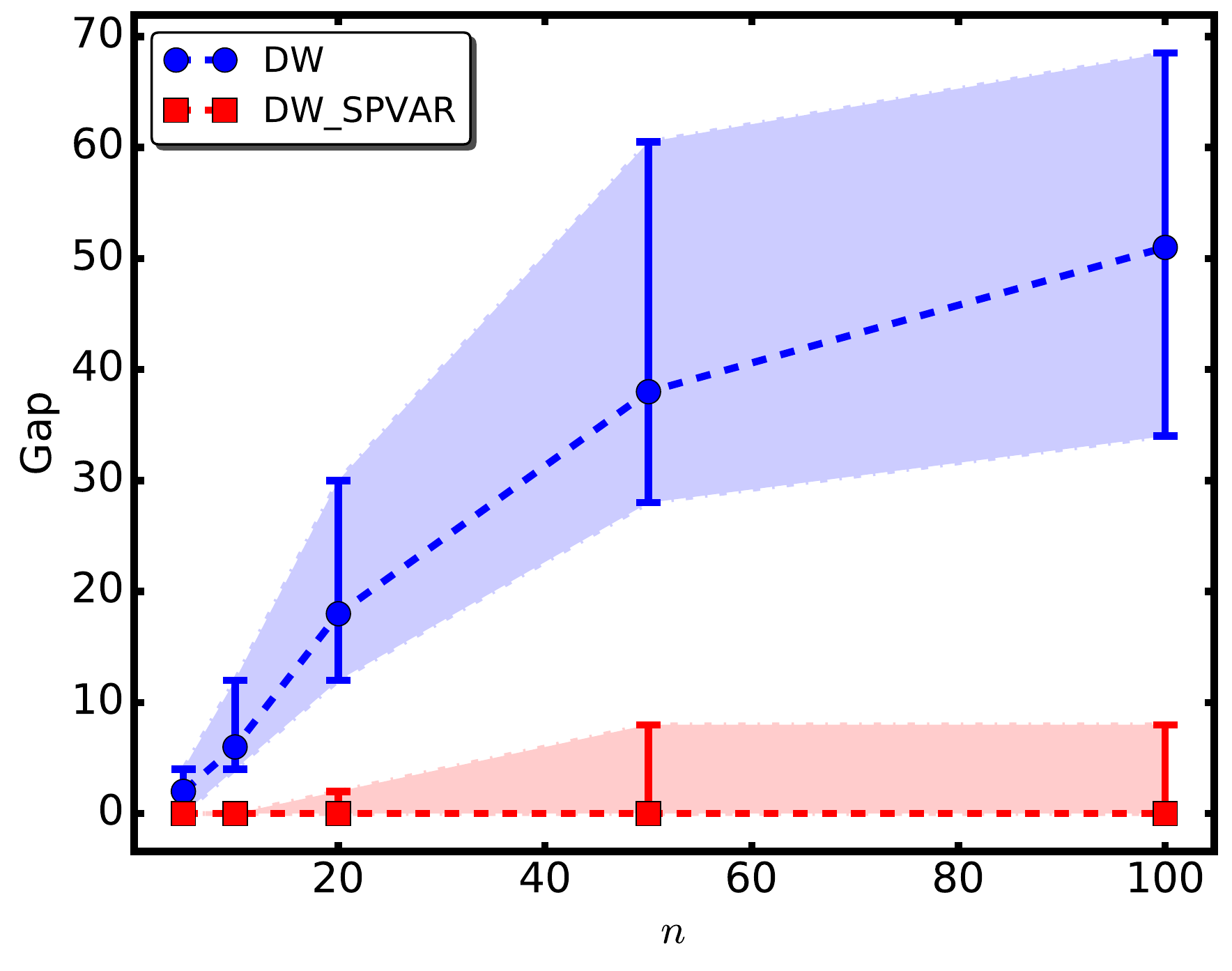} 
         \label{fig:chimera_gap}  
	}
     \quad
      \subfloat[$h=0$]{
         \includegraphics[width=0.45\textwidth]{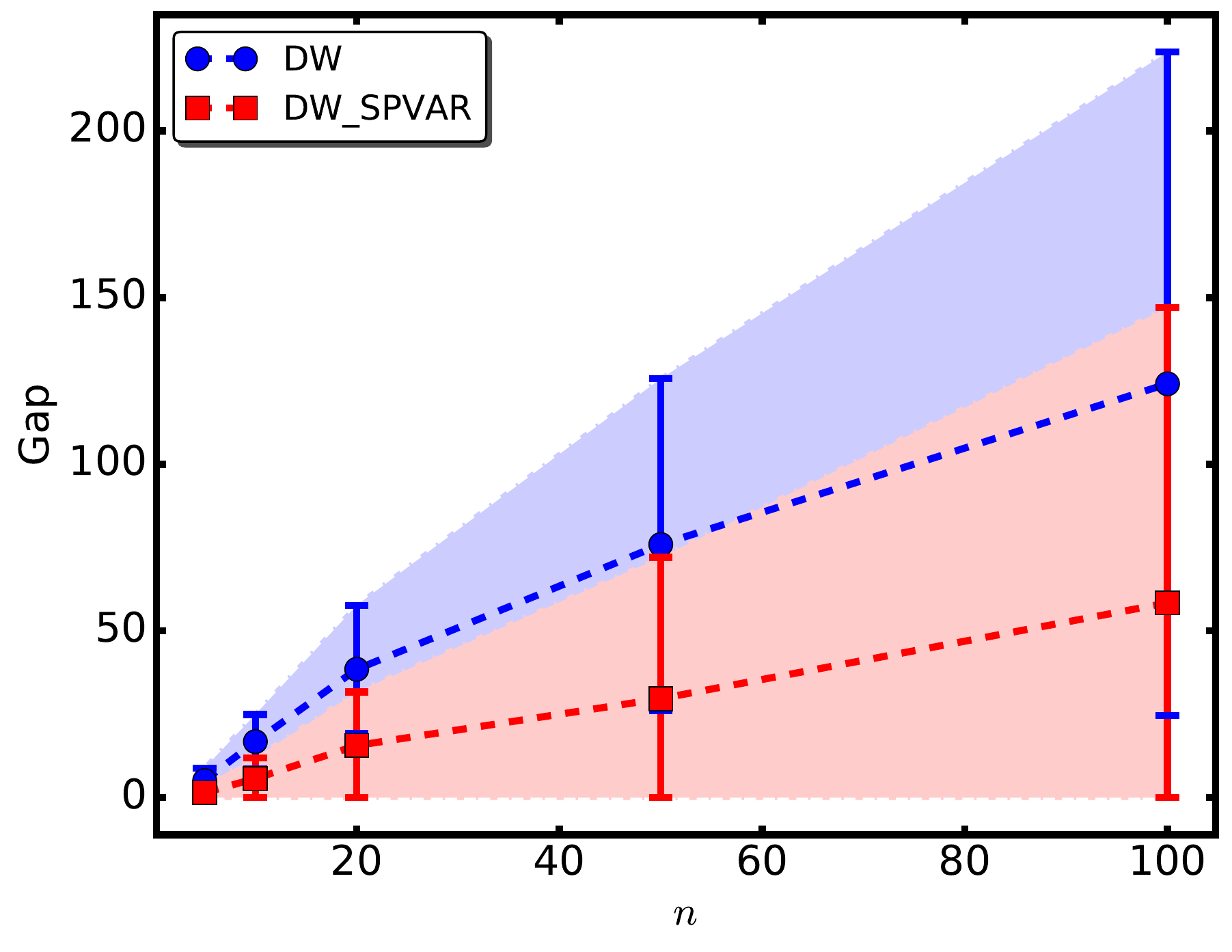} 
         \label{fig:chimera_h0_gap}  
	}
\caption{Success metrics for $U_{n,n-1,n-2}$ Chimera graph problems with reduced degeneracy. {\bf(a)} and {\bf(b)} The fraction of problems solved, as a function of $n$, for non-zero-bias problems and zero-bias problems, respectively. {\bf(c)} and {\bf(d)} The median gap, which is the energy difference between the best solution found and the best known solution, as a function of $n$, for non-zero-bias problems and zero bias-problems, respectively. In both figures, each point represents data from 100 random instances with the respective $n$.  }
\label{fig:chimera}
\end{figure*}

\begin{figure*}[!htbp]  
    \centering
    \subfloat[]{
        \includegraphics[width=0.45\textwidth]{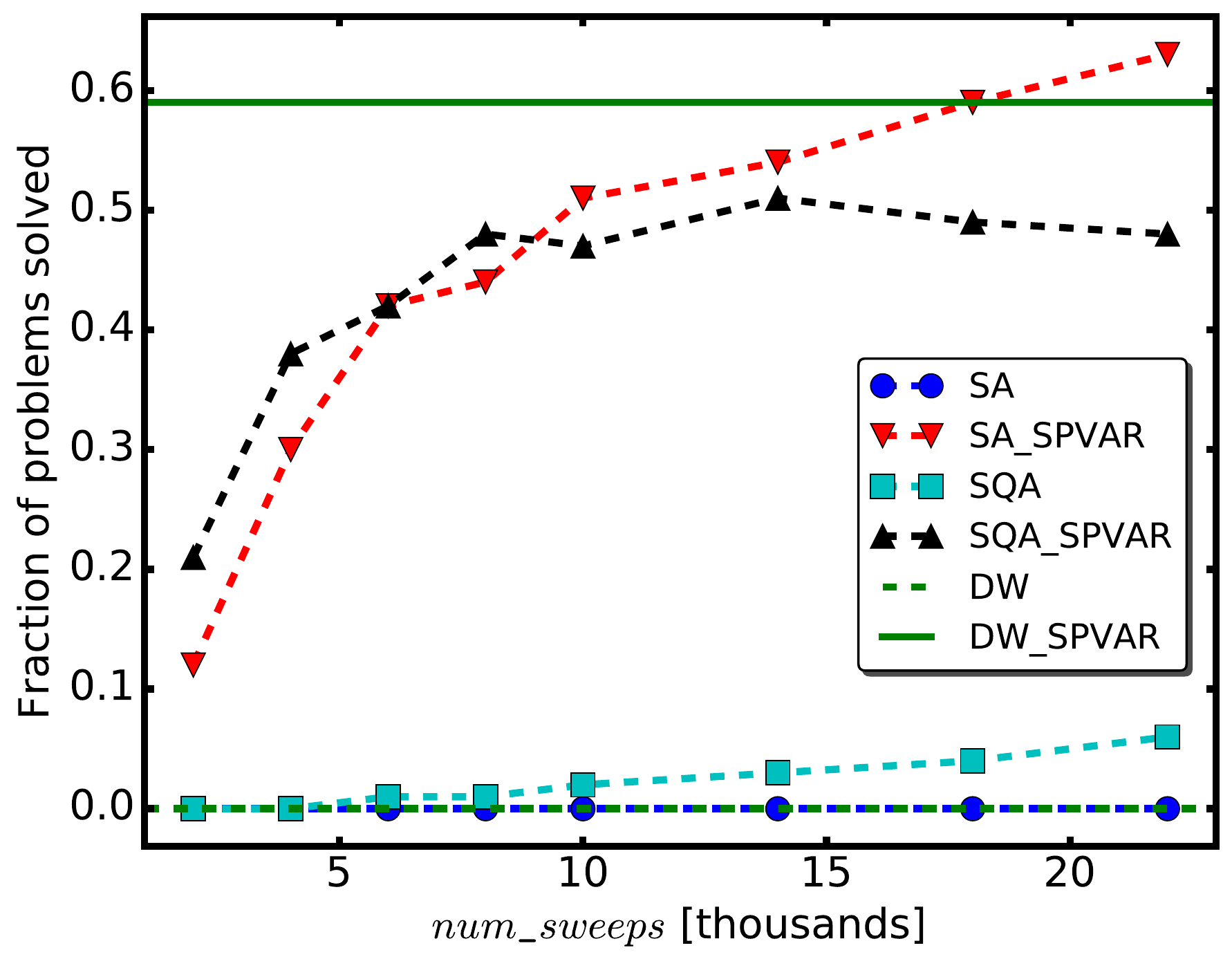} 
	\label{fig:chimera_vs_sweeps_fraction_problems_solved} 
	}
    \quad
   \subfloat[]{
         \includegraphics[width=0.45\textwidth]{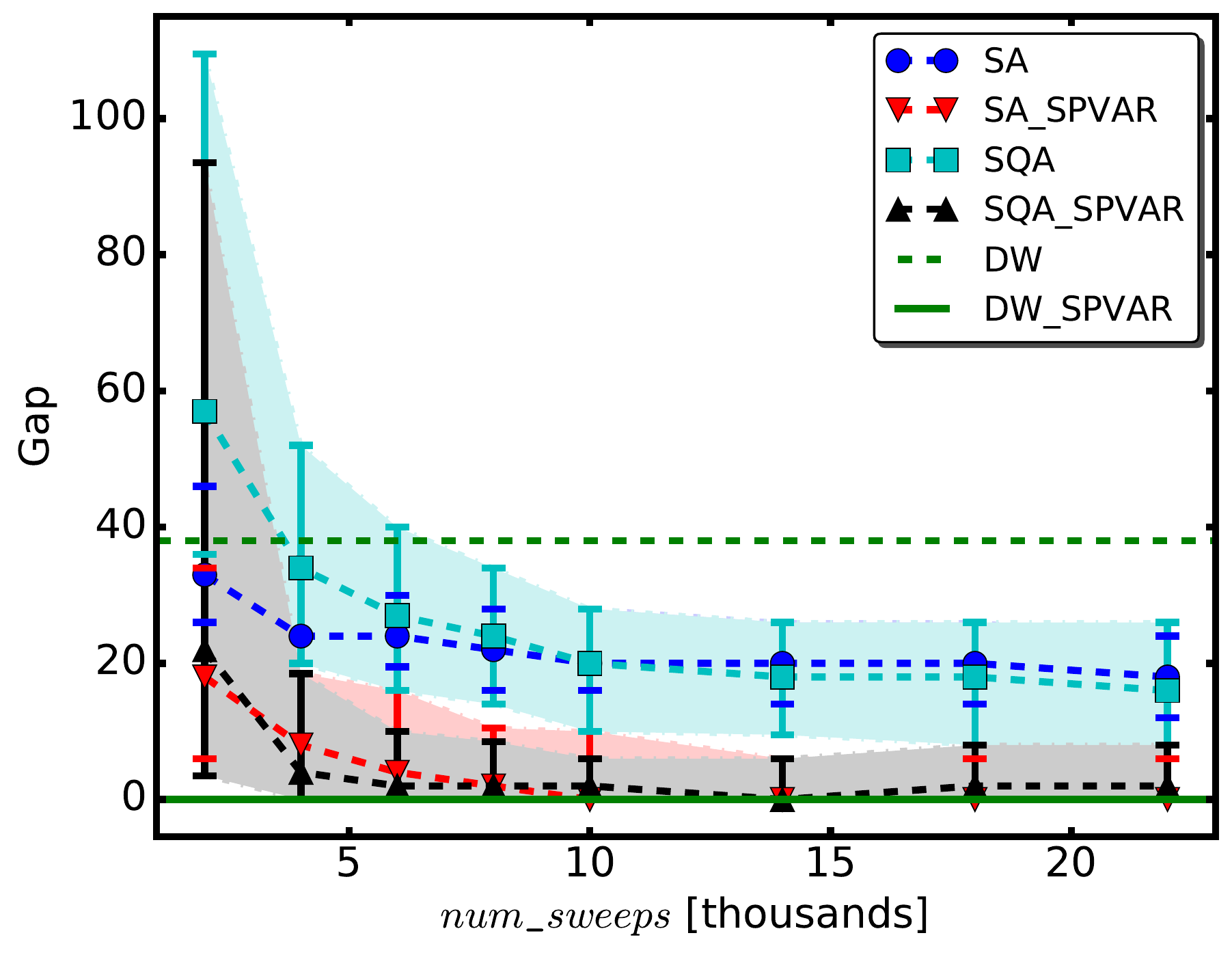} 
         \label{fig:chimera_vs_sweeps_gap}  
	}
\caption{Success metrics for $U_{50,49,48}$ Chimera graph problems with non-zero bias and reduced degeneracy, versus the number of sweeps. {\bf(a)} The fraction of problems solved. {\bf(b)} The median gap, which is the energy difference between the best solution found and the best known solution. The figures show median values for 100 random instances, solved with differing numbers of sweeps. Horizontal lines indicate the median values for D-Wave's quantum annealer, without and with Multi-start SPVAR (`DW' and `DW\textunderscore SPVAR', respectively).}
\label{fig:chimera_vs_sweeps}
\end{figure*}

We present results for the success metrics of $U_{n,n-1,n-2}$ Chimera graph problems with reduced degeneracy with non-zero and zero bias in Figure~\ref{fig:chimera} for the quantum annealer, with and without applying Multi-start SPVAR. We present results comparing success metrics for SA, SQA, and the quantum annealer (the DW2X\textunderscore SYS4) for $U_{50,49,48}$ problems with non-zero bias and reduced degeneracy in Figure~\ref{fig:chimera_vs_sweeps}. For the zero-bias problems, we set the \textit{elite\textunderscore threshold} adaptively as suggested in \cite{karimi2017boosting}. 


The results presented in Figure~\ref{fig:chimera} show that these problems are very hard for the quantum annealer, with a negligible success rate for $n>5$ for both the non-zero-bias and zero-bias problems. We attribute this difficulty to ICE, given that the accuracy required to distinguish different couplers scales with $1/n$ for these problems. In addition, the reduction of the ground state degeneracy by eliminating local degeneracies is likely a contributing factor. Zero-bias problems are known to be harder, in general, and indeed for $n=5$, the results obtained from the quantum annealer are better for the non-zero-bias problems.

Despite the very low success rate of the quantum annealer used alone, when coupled with our method, the quantum annealer solved over $60\%$ of each of the problem sets for the non-zero-bias problems, and close to $20\%$ for the zero-bias problems. As noted in \cite{karimi2017boosting}, our intuition is that the quantum annealer is better at finding low-energy states than it is at finding optima. We hypothesize that when considering only the state with the lowest energy value in the sample, additional information about the low-energy landscape contained within the rest of the sample is discarded. Utilizing this information allows our method to substantially improve the success metrics. 

Figure~\ref{fig:chimera_vs_sweeps} shows that the success metrics of SA and SQA improve as the number of sweeps is increased, albeit with diminishing returns. The results also show that both SA and SQA are able to match the quantum annealer's performance using a relatively modest 2000 sweeps. However, if our method is applied to all samplers, SA and SQA require a significantly larger number of sweeps to match the performance of the quantum annealer when using our method. Therefore, in this case, the quantum annealer seems to benefit more from the application of our method. It is worth mentioning that, combined with SPVAR, QA was able to solve 60$\%$ of the problems roughly three orders of magnitude faster than either of SA or SQA combined with SPVAR. The median R99 for SA was 43,709, and for QA it was 66,745, and the time required to obtain a single solution was 20\,ms for SA (for this problem size---1100 variables) and 20\,$\mu$s for QA (the annealing time). 

\subsubsection{Scaling analysis}

\begin{figure}[!htbp]  
 \subfloat[]{
     \centering
    \includegraphics[width=0.5\textwidth]{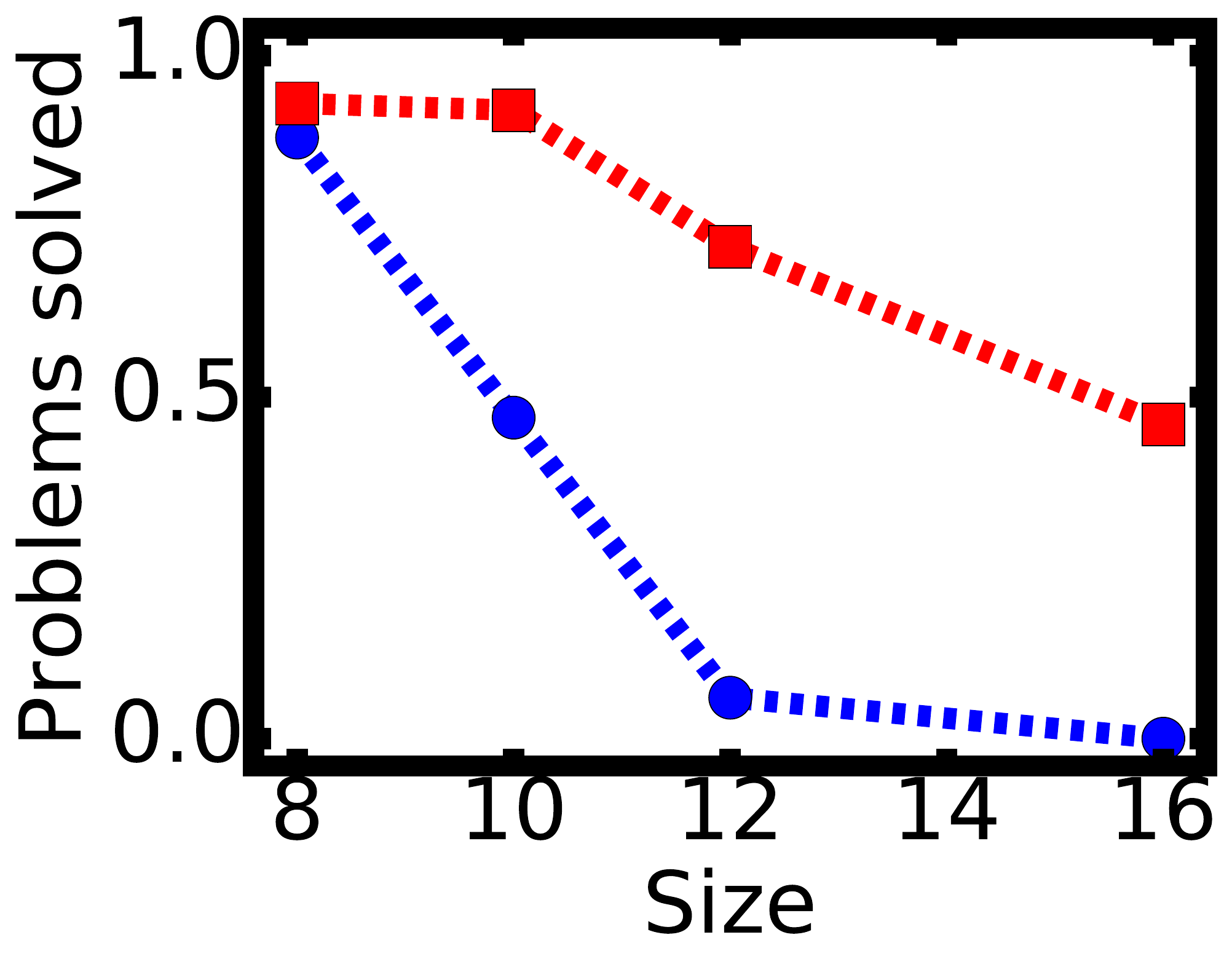} 
    \label{fig:chimera_vs_size_fraction_problems_solved} 
 	}
  \subfloat[]{
    \centering
    \includegraphics[width=0.5\textwidth]{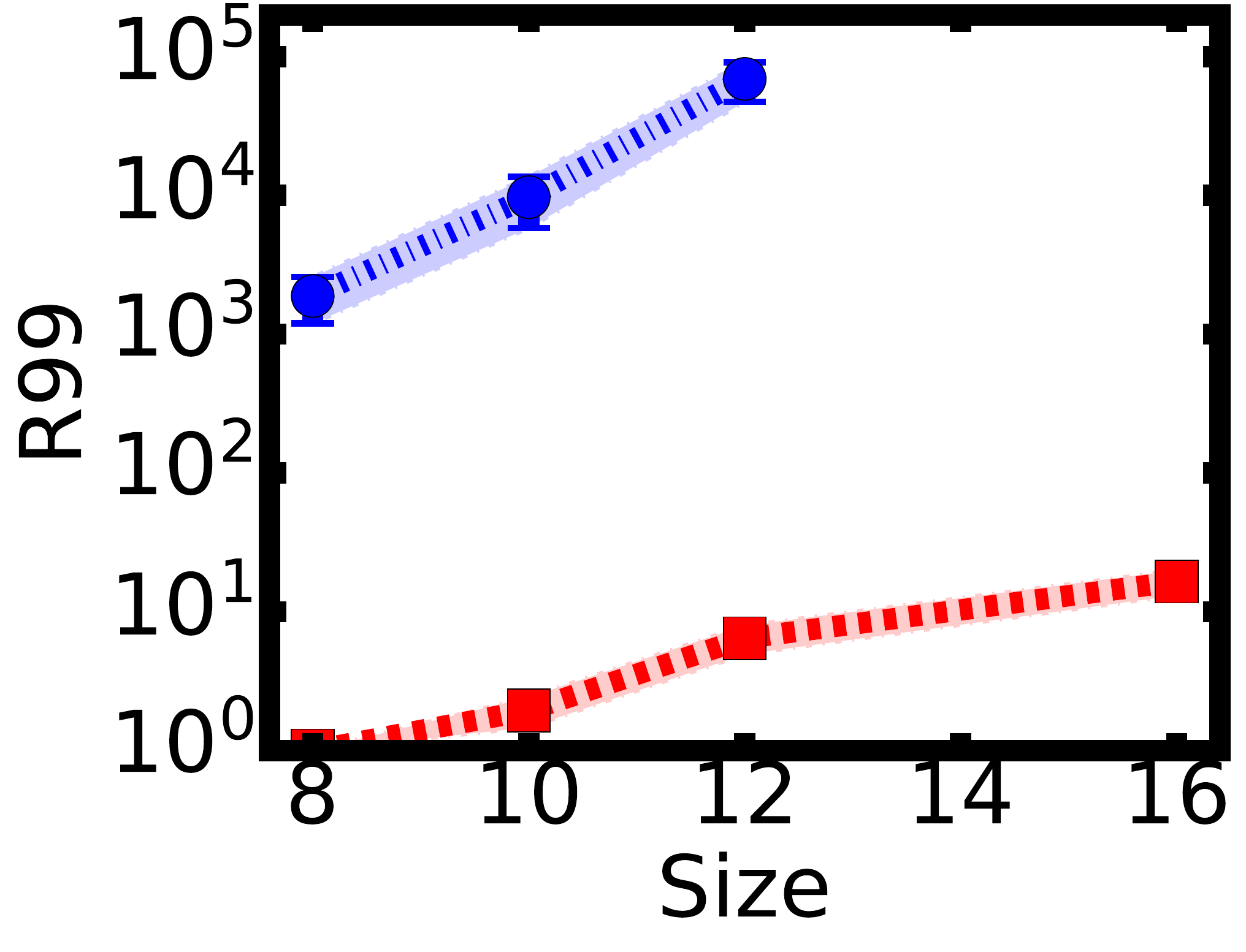} 
    \label{fig:chimera_vs_size_R99} 
    } \\
    \vspace{-0.2cm}
    \subfloat[]{
    \centering
	\includegraphics[width=0.5\textwidth]{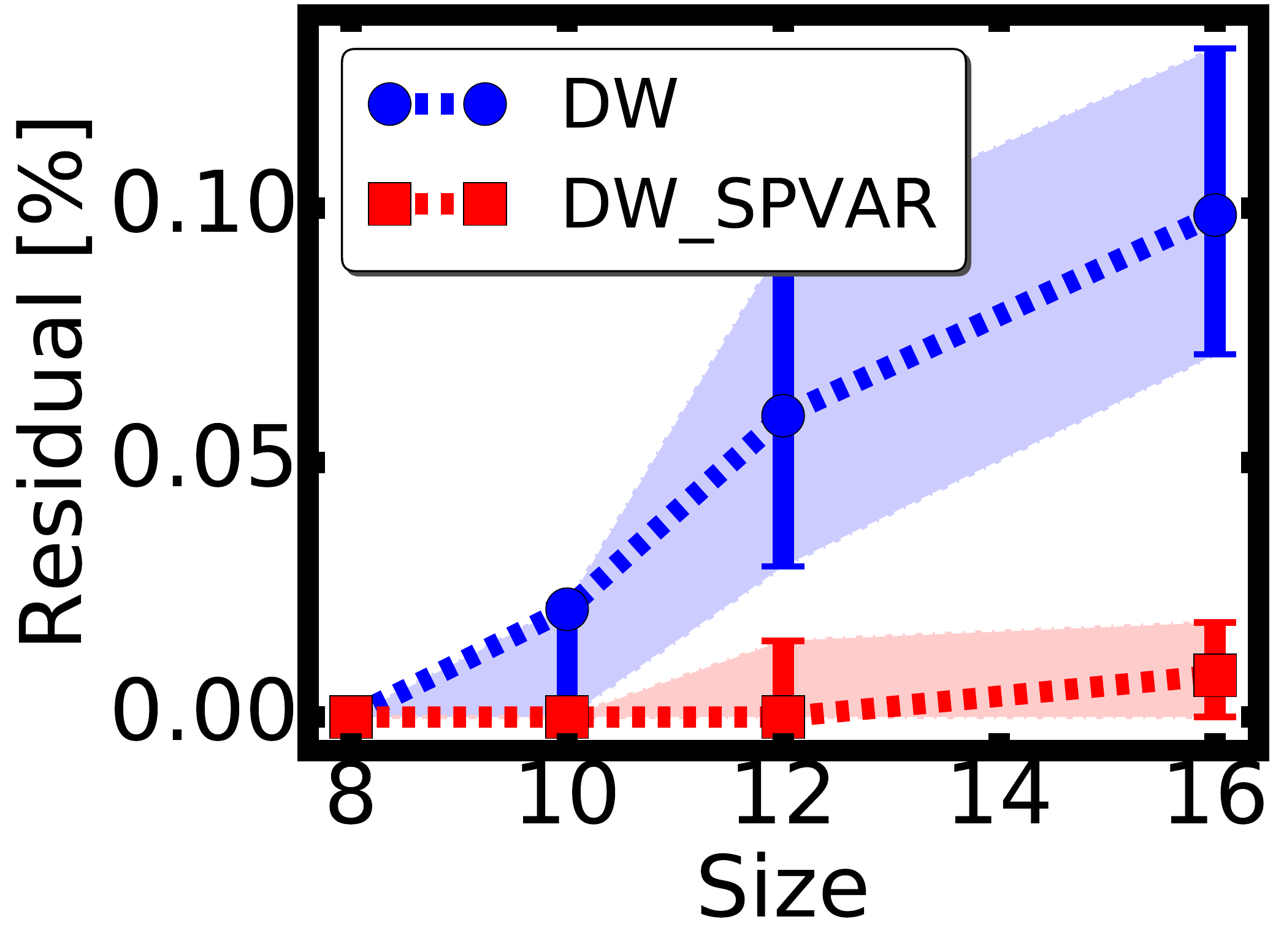} 
    \label{fig:chimera_vs_size_gap} 
 }
  \subfloat[]{
    \centering
    \includegraphics[width=0.5\textwidth]{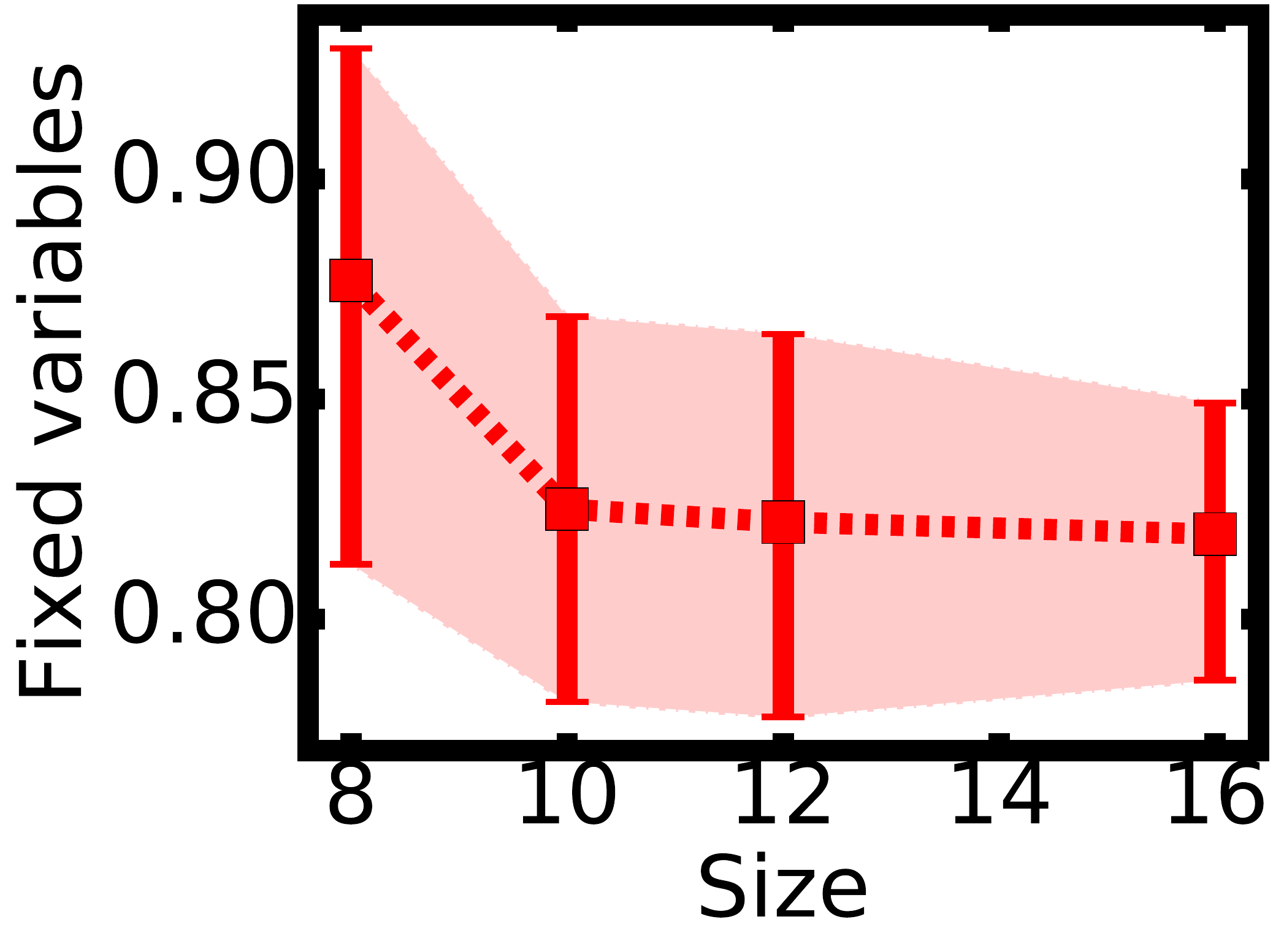} 
    \label{fig:chimera_vs_size_num_fixed_qubits} 
}
  \caption{Success metrics for Multi-start SPVAR on Chimera graph-structured problems in $U_{10}$, versus size. \textbf{(a)} The fraction of problems solved at each size. \textbf{(b)} The median R99 values, calculated using bootstrapping, with and without using SPVAR. \textbf{(c)} The median energy residual. \textbf{(d)} The median fraction of fixed variables at each size.}
  \label{fig:chimera_vs_size}
\end{figure}

In this section, we investigate the scaling of success metrics when using SPVAR. We generated four native Chimera graph problem sets from ${U_{10} \in \{-10, \dots, 10\}}$ (0 was excluded for the couplers but included for the biases), each consisting of 100 problems with increasing sizes, from a Chimera graph of size 8 to 16 (the latter referring to a 16 $\times$ 16 grid of blocks). The quantum annealer we used for this study was the \mbox{D-Wave 2000Q}, which has 2048 qubits, with a qubit yield of almost $99\%$ \footnote{\protect The chip at our disposal has 2023 active qubits, a working temperature of 14 $\pm$1 mK, and a minimum annealing time of 5 $\mu$s.}. In Figure~\ref{fig:chimera_vs_size}, we show the dependence of success metrics on the problem size.

Figure~\ref{fig:chimera_vs_size_fraction_problems_solved} shows that the fraction of problems solved decreases exponentially for the quantum annealer used alone, but scales better when used in conjunction with \mbox{SPVAR}. Note that for size 16, none of the problems were solved using the quantum annealer alone, but $46\%$ of problems were solved when using the quantum annealer in conjunction with SPVAR. 

We also observed significant improvement in the R99, which is a measure of the time to solution \footnote{\protect{ The R99 is the number of annealing cycles required to find the best known energy with a confidence of 99\%.}}, shown in Figure~\ref{fig:chimera_vs_size_R99}. The reported R99 value is the mean of the median R99 in 1000 bootstrapped samples (using only the R99s that could be measured). In cases in which less than 50\% of the problems were solved, this value is a lower bound of the actual median R99. In particular, for size 12, the quantum annealer solved only 6 problems, so the R99 value is most likely a very loose lower bound. The scaling when using SPVAR is clearly improved. 

Figure~\ref{fig:chimera_vs_size_gap} shows a clear advantage in scaling when our method is utilized. In Figure~\ref{fig:chimera_vs_size_num_fixed_qubits}, the number of variables fixed appears to plateau, such that even at large sizes the method fixes more than 80\% of the variables. 

\subsection{3D spin glass problems}
\label{sec:results_3Dspin}

\begin{figure*}[!htbp]  
    \centering
    \subfloat[]{
        \includegraphics[width=0.45\textwidth]{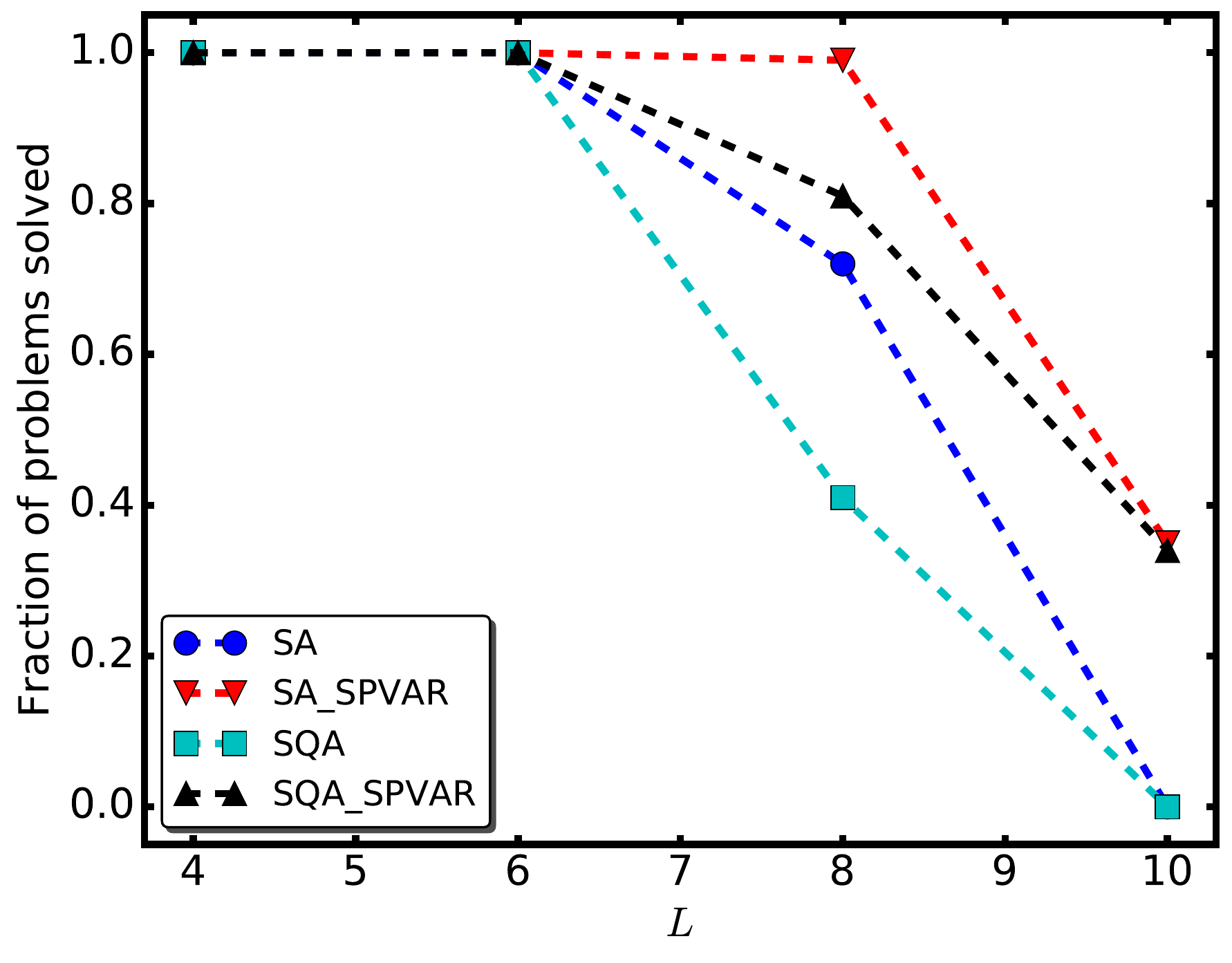} 
	\label{fig:gs_vs_L_fraction_problems_solved} 
   }
    \quad
    \subfloat[]{
         \includegraphics[width=0.45\textwidth]{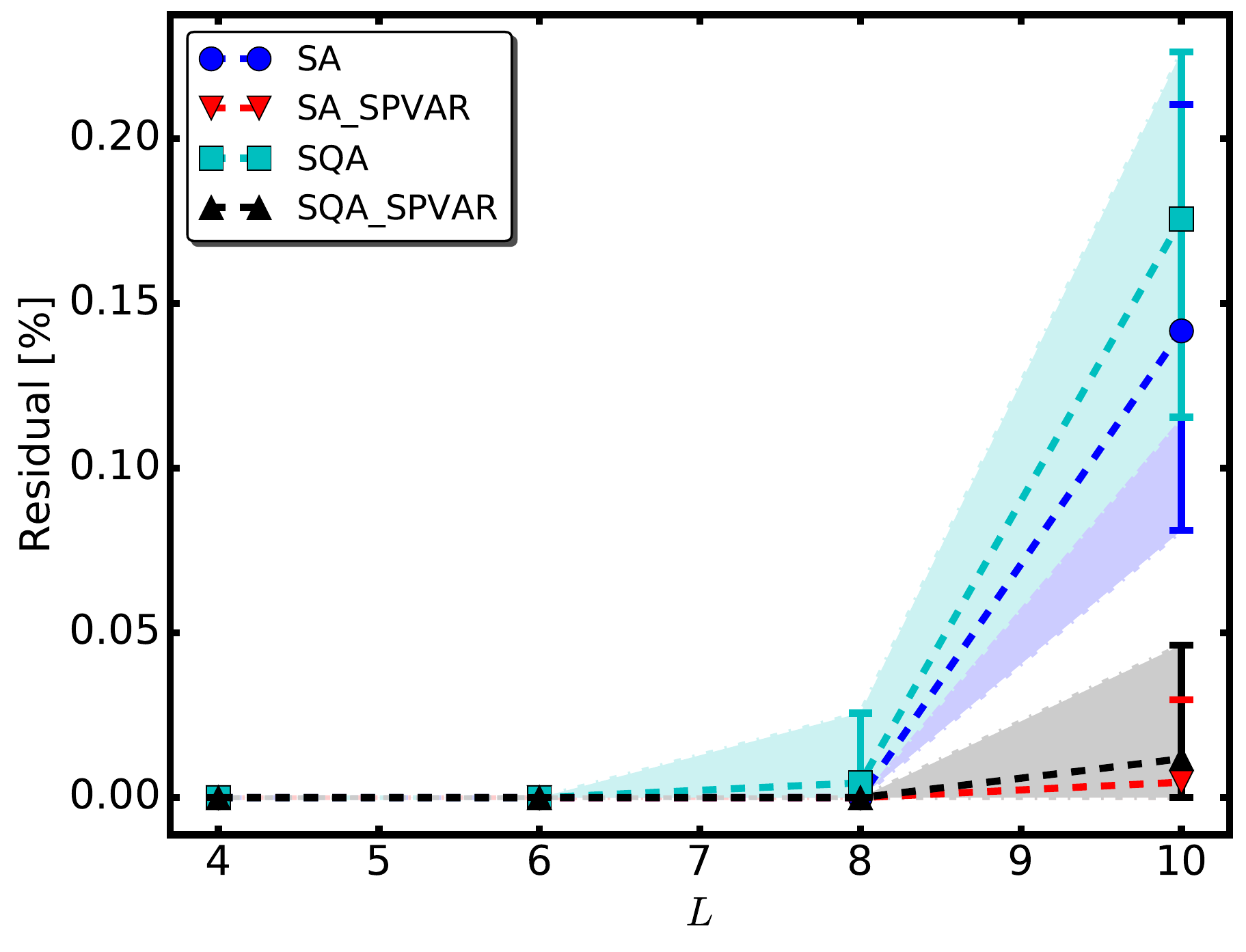} 
         \label{fig:gs_vs_L_residual}  
	}
\caption{Success metrics for 3D spin glass problems, versus size. {\bf(a)} The fraction of problems solved, as a function of $L$, where the size is $L^3$. {\bf(b)} The median residual, which is the relative energy difference (in percent) between the best solution found and the best known solution, as a function of $L$. In both figures, each point represents data from 100 random instances with the respective $L$. }
\label{fig:gs_vs_L}
\end{figure*}

\begin{figure*}[!htbp]  
    \centering
   \subfloat[$L=8$]{
            \includegraphics[width=0.45\textwidth]{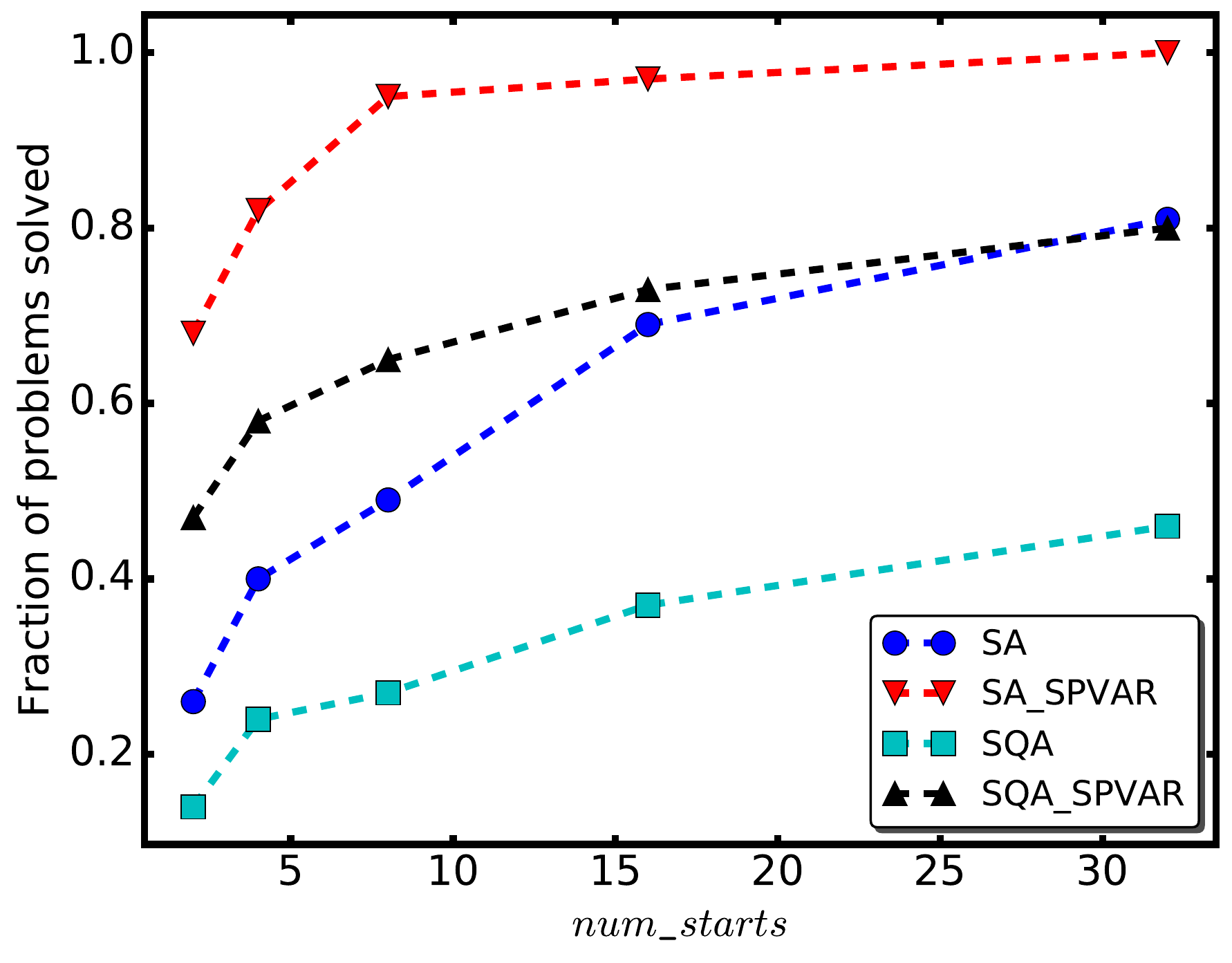} 
      \label{fig:gs_vs_calls_L8_fraction_problems_solved}
	}
      \quad
	\subfloat[$L=10$]{
            \includegraphics[width=0.45\textwidth]{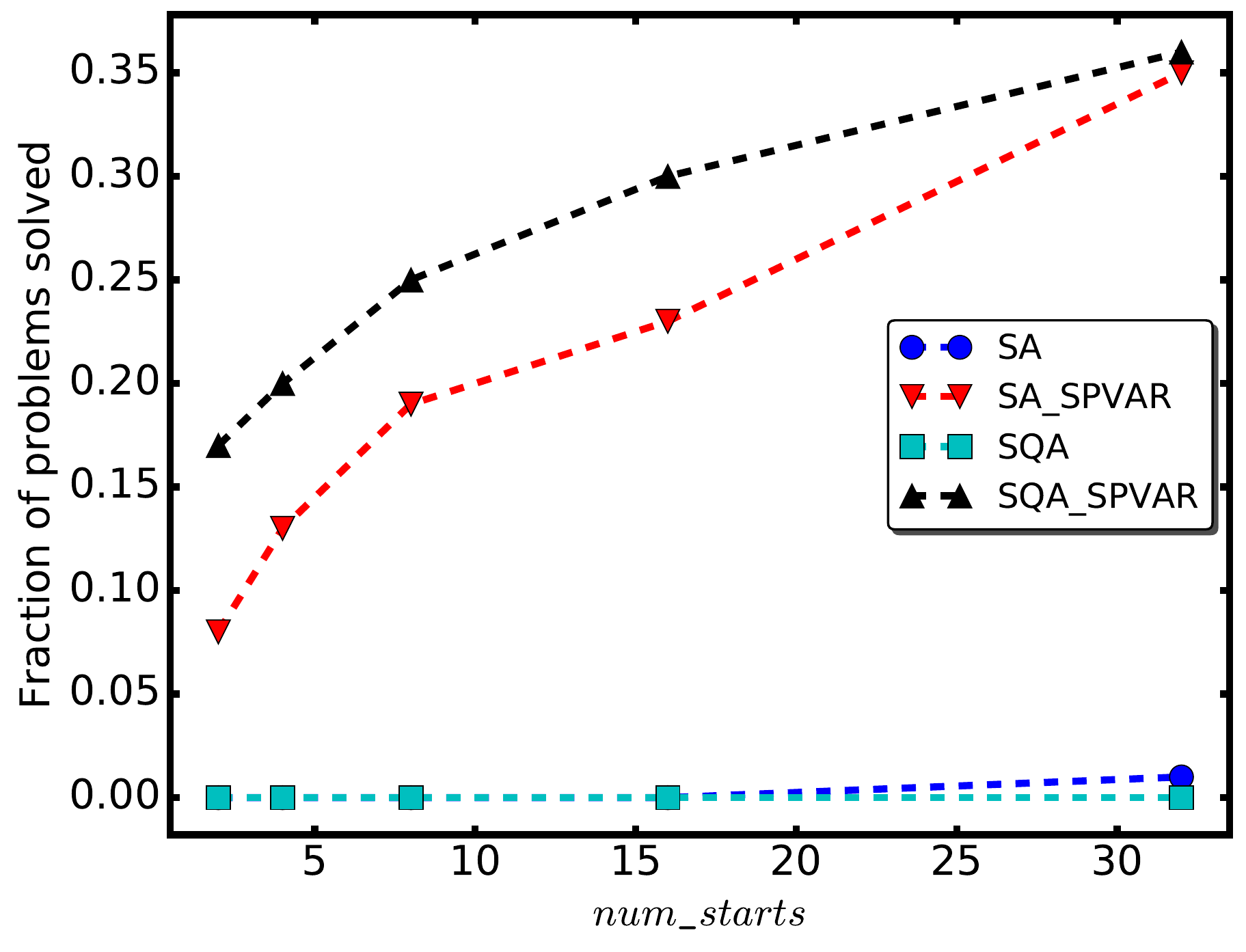} 
      \label{fig:gs_vs_calls_L10_fraction_problems_solved}
	} \\
	\subfloat[$L=8$]{
        \includegraphics[width=0.45\textwidth]{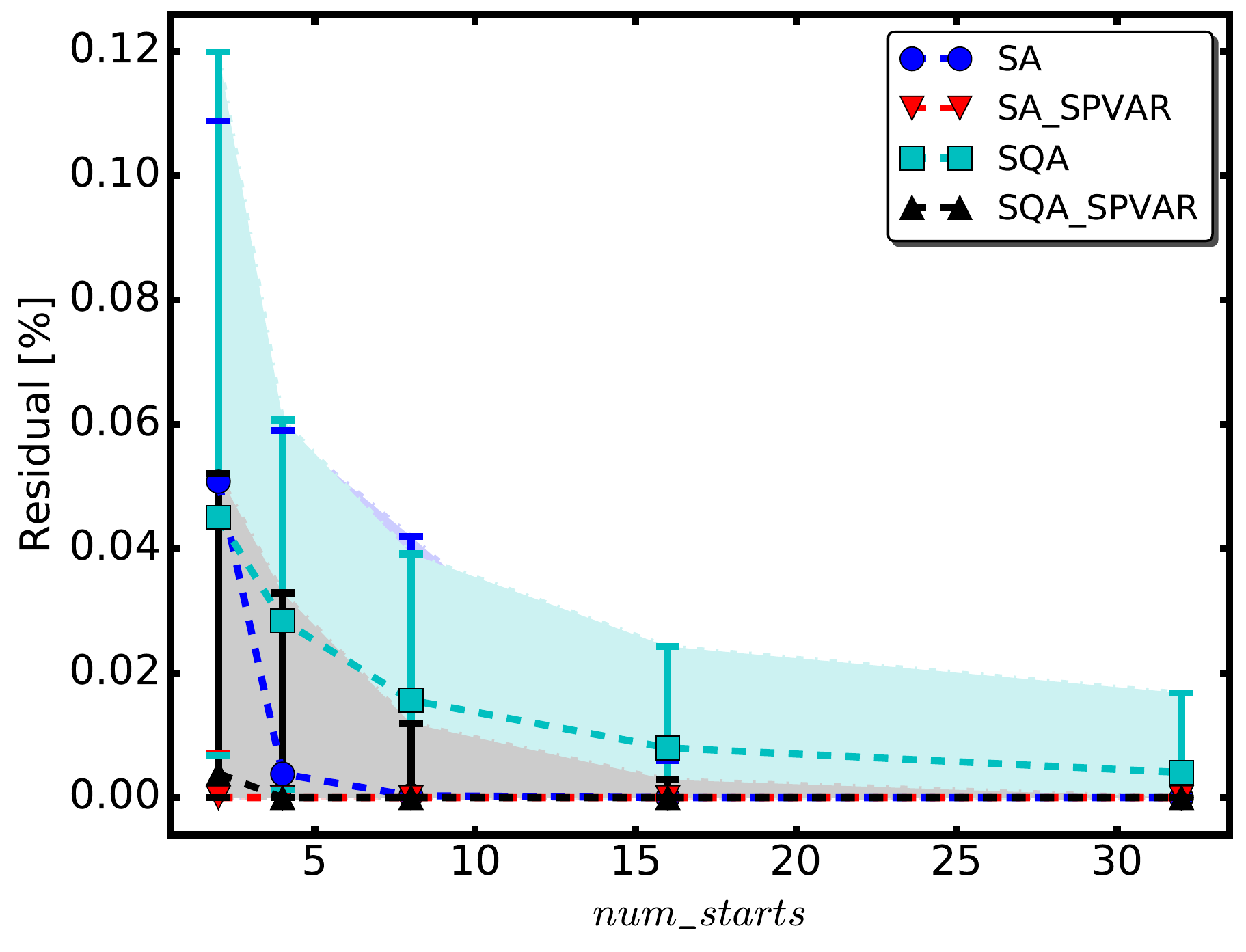} 
        \label{fig:gs_vs_calls_L8_residual}
	}
    \quad
	\subfloat[$L=10$]{
        \includegraphics[width=0.45\textwidth]{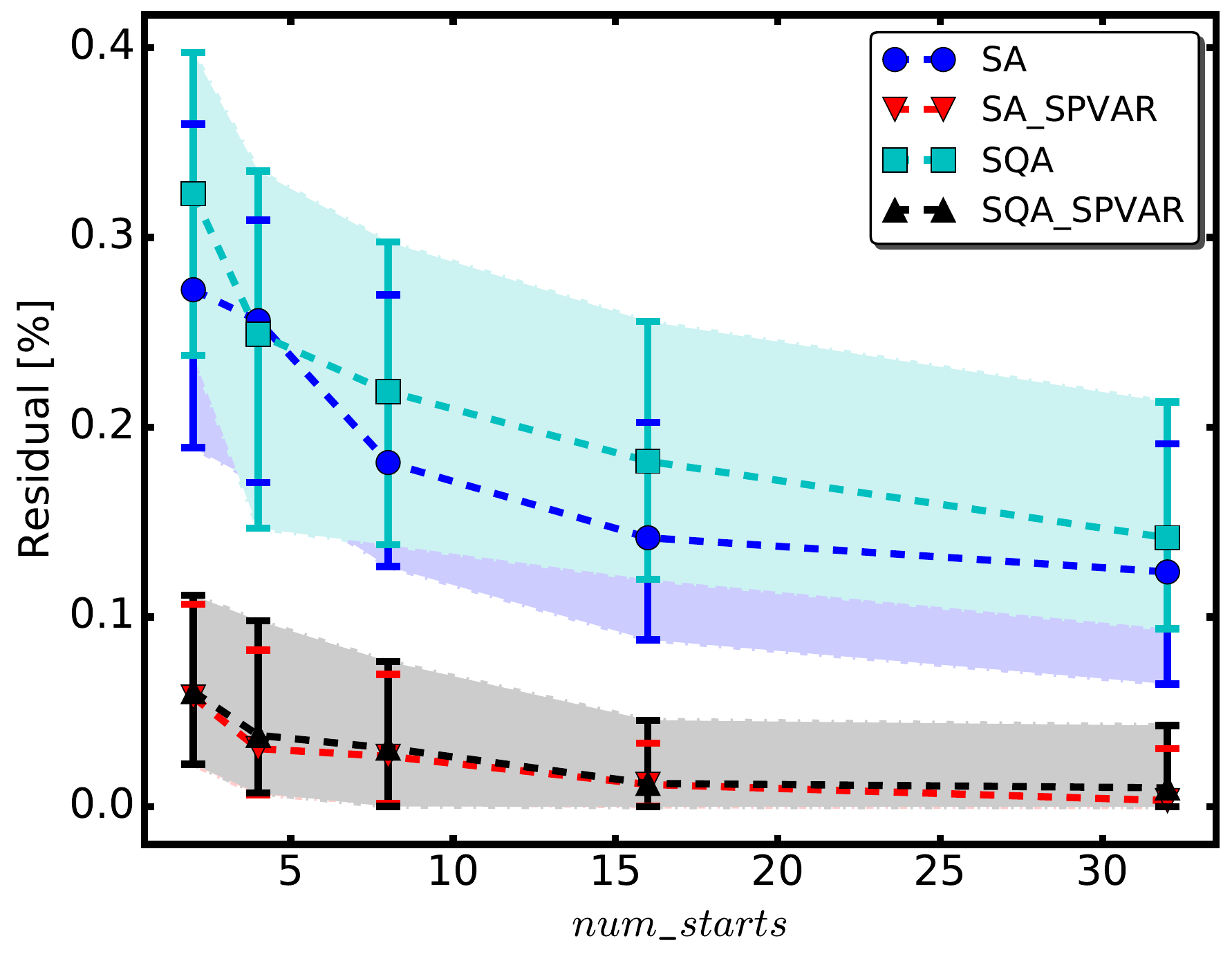} 
        \label{fig:gs_vs_calls_L10_residual}
	}
\caption{Success metrics for 3D spin glass problems, versus $num\textunderscore starts$. {\bf(a)} and {\bf(b)} The fraction of problems solved, for $L=8$ and $L=10$ (respectively). {\bf(c)} and {\bf(d)} The median residual, which is the relative energy difference (in percent) between the best solution found and the best known solution, for $L=8$ and $L=10$ (respectively). In all figures, each point represents data from 100 random instances.}
\label{fig:gs_vs_calls}
\end{figure*}

Ising problem instances generated on a 3D cubic lattice have been shown to be relatively hard instances, for example, by benchmarking with parallel tempering and population annealing \cite{wang2015comparing,wang2015population}. We obtained and benchmarked a selection of instances with \mbox{$L=4,6,8,10$} from the authors of those studies, where the number of variables is $L^3$. For each $L$, we chose the 20 hardest instances from a larger instance pool and selected an additional 80 instances uniformly randomly from that same pool. We present results for the success metrics as a function of $L$ for different samplers in Figure~\ref{fig:gs_vs_L}, and results for the dependence of the success metrics on $num\textunderscore starts$ in Figure~\ref{fig:gs_vs_calls}. As these problems could not be embedded on the quantum annealer chip to which we had access, we present the results of only SA and SQA for these problems.


In Figure~\ref{fig:gs_vs_L}, we see that the application of our method provides a substantial improvement in the success metrics, especially for the larger and harder problem instances. For example, for $L=10$, no problems are solved by SA and SQA, but with the addition of our method, they are both able to solve almost 40\% of the problems, with a greatly reduced residual. Increasing \textit{num\textunderscore starts}, as in Figure~\ref{fig:gs_vs_calls}, improves the results of all methods, albeit with diminishing returns. For the largest problems ($L=10$), increasing \textit{num\textunderscore starts} barely improves the results for the samplers when our method is not applied, but gives a substantial improvement when our method is applied. The performance is comparable with population annealing and parallel tempering \cite{wang2015comparing}.

\subsubsection{Parallel tempering with isoenergetic cluster moves}
\label{sec:results_pticm}

For all solvers, except for the parallel tempering with isoenergetic cluster moves algorithm (PTICM), the sample was formed by multi-starting the given algorithm and aggregating the results. This is a natural choice for algorithms that return a single state (or a finite number of states) per start. PTICM returns a single state per replica. However, forming the sample in this way suffers from the disadvantage that each replica might have been in many lower-energy states during the optimization process, which would be ignored. In addition, since the temperatures of the replicas typically span a large range, many of the replicas return relatively high-energy states, such that the sample formed would not be a good low-energy sample. For this reason, we chose instead to keep track of the 10 lowest-energy states found for each of the replicas in the lower half of the temperature range.

The results for PTICM are presented separately from the other solvers. One reason for this is that the sample was formed in an inherently different way; for the other solvers, the sample consisted of independently sampled states. In addition, there is no obvious way to equate the computational effort used by PTICM to that of SA or SQA, since a single sweep is a combination of a Metropolis update, a parallel tempering move, and an isoenergetic cluster move. Finally, the performance of PTICM was superior on the instances benchmarked with the other solvers, leaving little room for improvement by SPVAR. For example, the 3D spin glass problems of sizes $L \leq 8$ were all solved by PTICM without SPVAR, and even for $L=10$, most were solved.

\begin{figure*}[!htbp]  
    \centering
    \subfloat[]{
            \includegraphics[width=0.45\textwidth]{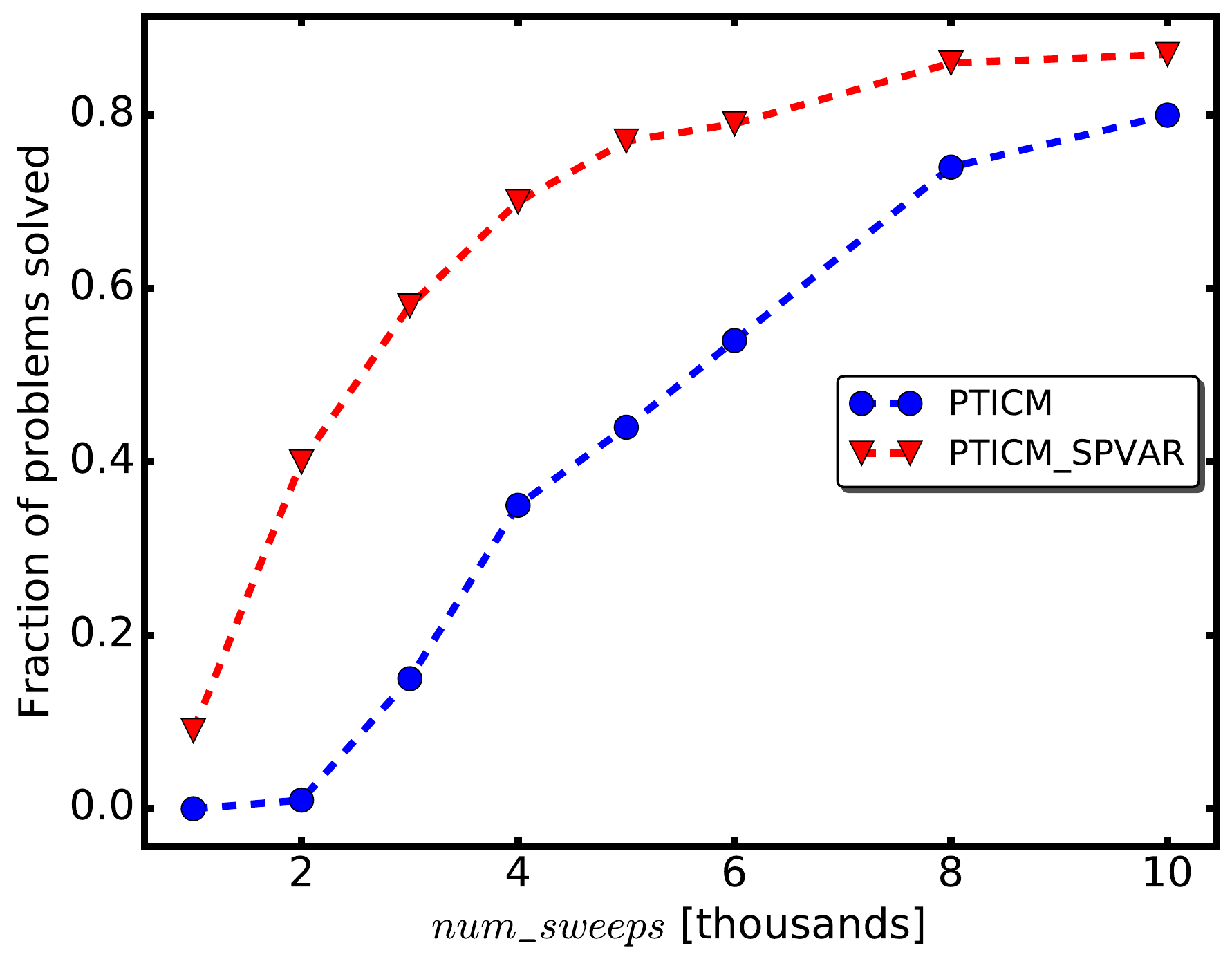} 
      \label{fig:gs_PTICM_vs_sweeps_not_planted_L10_fraction_problems_solved}
	}
      \quad
	\subfloat[]{
            \includegraphics[width=0.45\textwidth]{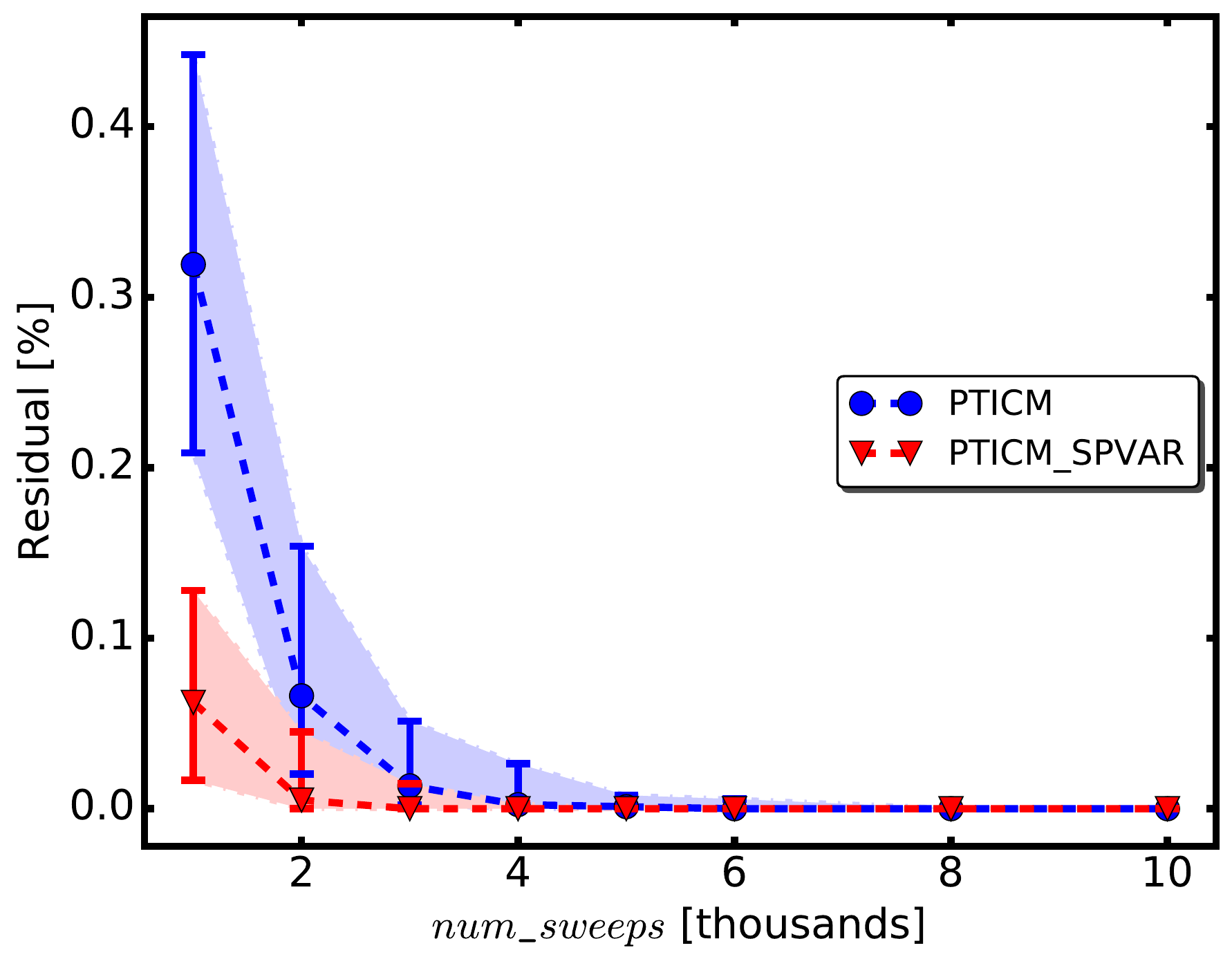} 
      \label{fig:gs_PTICM_vs_L_sweeps10000_combined}
    }
\caption{Success metrics for PTICM for 3D spin glass problems with ${L=10}$, versus \textit{num\textunderscore sweeps}. {\bf(a)} The fraction of problems solved. {\bf(b)} The median residual, which is the relative energy difference (in percent) between the best solution found and the best known solution. In both figures, each point represents data from 100 random instances.}
\label{fig:gs_PTICM_vs_sweeps_not_planted_L10}
\end{figure*}

\begin{figure*}[!htbp]  
    \centering
   \subfloat[]{
            \includegraphics[width=0.45\textwidth]{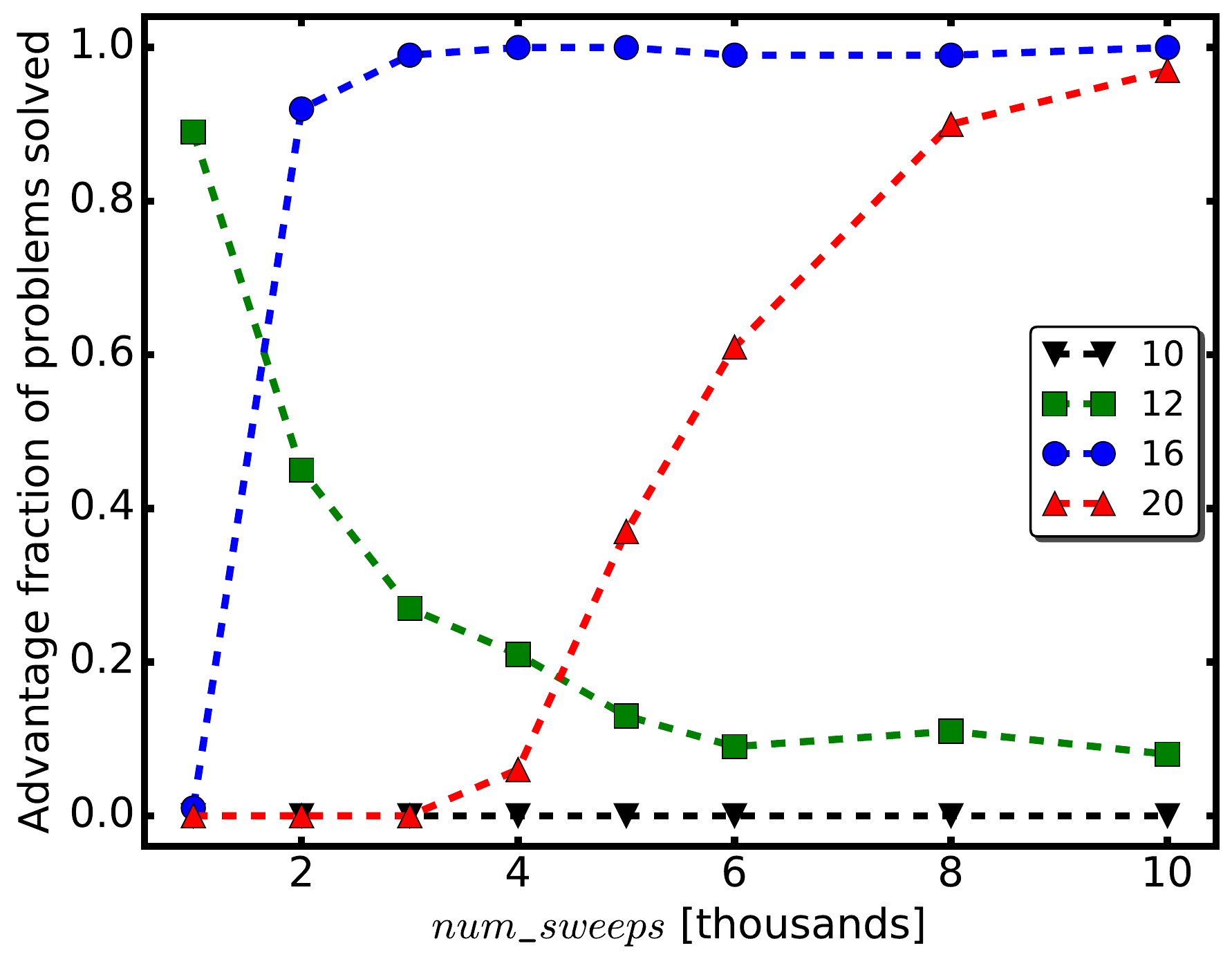} 
      \label{fig:gs_PTICM_vs_sweeps}
	}
      \quad
	\subfloat[]{
            \includegraphics[width=0.51\textwidth]{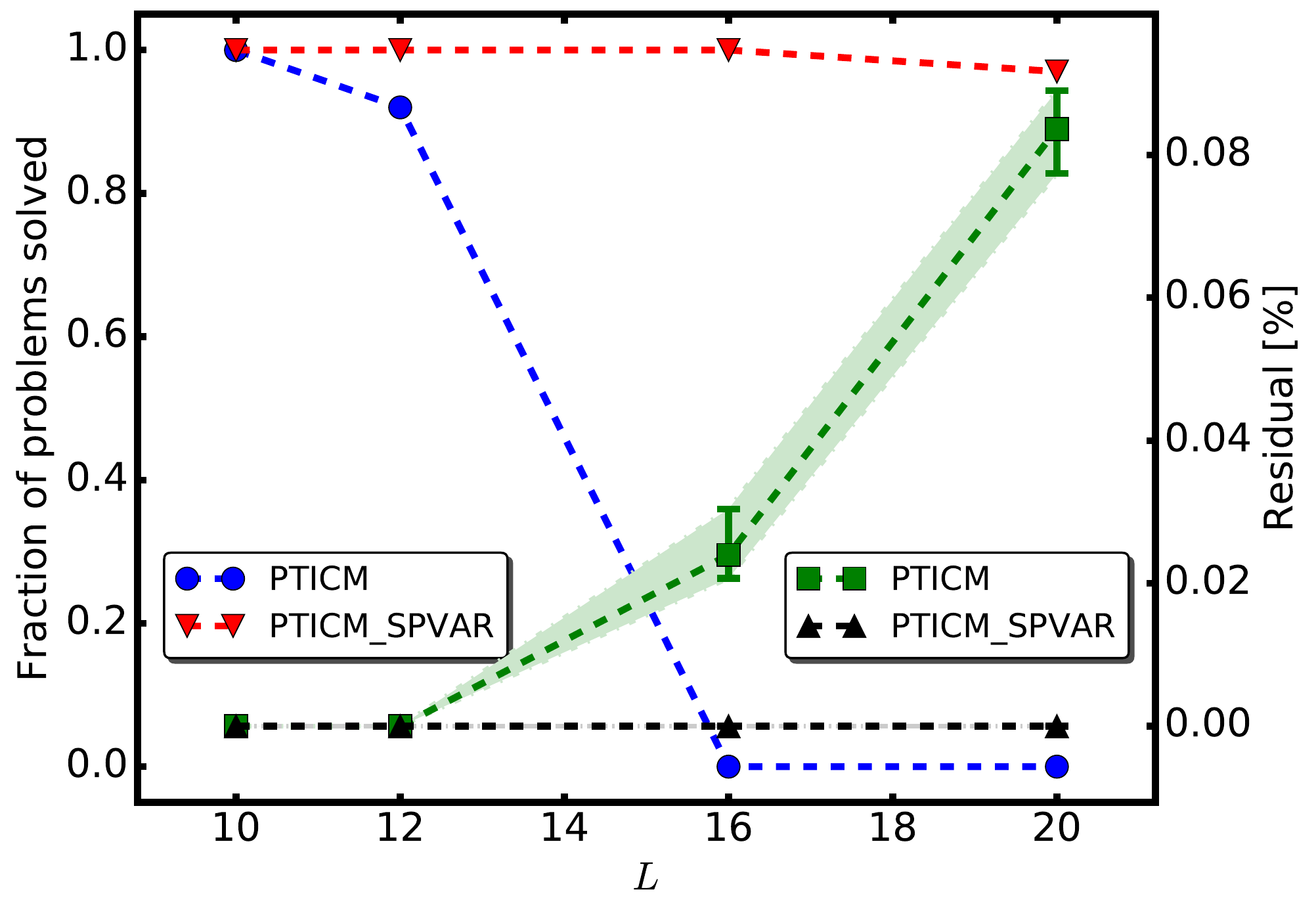} 
      \label{fig:gs_PTICM_vs_L}
   }
\caption{Fraction of problems solved for 3D spin-glass problems with planted solutions, versus num sweeps and $L$. Panel {\bf (a)} shows the difference in the fraction of problems solved for $L = 10\text{--}20$ between PT+ICM with SPVAR and PT+ICM without SPVAR. Note that for $L = 10$, PT-ICM solved all problems both with and without SPVAR. Panel {\bf (b)} shows the fraction of problems solved, and the median residual, which is the relative energy difference (in percent) between the best solution found and the best known solution, as a function of $L$. In both panels, each point represents data from 100 random instances with the respective $L$, solved with \textit{num\textunderscore sweeps} = $10^4$.}
\end{figure*}

The results for the same pool of $L=10$ problems from Section~\ref{sec:results_3Dspin} are presented in Figure~\ref{fig:gs_PTICM_vs_sweeps_not_planted_L10}. For $L \leq 8$, all instances were solved by PTICM as well as by PTICM with SPVAR, so we do not present results for those problems. In order to probe the performance of PTICM on larger problems and to study the scaling, we procured 3D spin glass instances with $L=10,12,16,20$ with planted solutions \cite{wang2017spin}. This was necessary, as for the larger problems, a huge computational effort is required to ensure that the ground state is found with high confidence using heuristics. We present results for the success metrics as a function of $num\textunderscore sweeps$ in Figure~\ref{fig:gs_PTICM_vs_sweeps}, and results for the dependence of the success metrics on $L$ in Figure~\ref{fig:gs_PTICM_vs_L}. 

In Figure~\ref{fig:gs_PTICM_vs_sweeps_not_planted_L10}, we see that our method provides a modest improvement for the non-planted 3D spin glass problems of size $L=10$. In Figure~\ref{fig:gs_PTICM_vs_sweeps}, we see that for $L \geq 12$, for almost all values of $num\textunderscore sweeps$, the fraction of problems solved is substantially larger. 
In Figure~\ref{fig:gs_PTICM_vs_L}, for the planted 3D spin glass problems, we see that our method provides a substantial improvement in both the success metrics and the scaling. For example, for $L=16$ and $L=20$, no problems are solved by PTICM, but with the addition of our method, it is able to solve almost all of the problems, with $num\textunderscore sweeps=10^4$. 
Interestingly, the $L=16$ and $L=20$ planted problems are harder for PTICM than the $L=10$ non-planted problems, but for PTICM with SPVAR this is reversed---it solved all of the $L=16$ planted problems, and almost all of the $L=20$ planted problems. We suspect that SPVAR is able to exploit the structure of the planted-solution problems.

\subsection{Fault diagnosis problems}
\label{sec:results_fault_diagnosis}

Fault diagnosis problems are one of the leading candidates for benchmarking quantum annealers~\cite{perdomo2015faultdetection, bian2015constrained,perdomo2017readiness}. We obtained a single problem set of 100 instances from 4-bit $\times$ 4-bit multiplier circuits as generated in Ref.~\cite{perdomo2017readiness}, where it was shown that these instances are at least one order of magnitude harder than conventionally used random spin glass instances. Besides their intrinsic hardness in terms of time to solution, these fault diagnosis instances are also harder from an asymptotic scaling perspective, making them an interesting testbed on which to apply SPVAR.

In order to solve these problems with D-Wave's quantum annealer, special treatment was required since their adjacency matrix differs from the hardware graph of the quantum annealer (a Chimera graph). A common method is to find an \emph{embedding}, which is a mapping from each logical variable to one or more qubits, referred to as a \emph{chain}. The identification of multiple qubits with a single logical variable results in a higher effective connectivity, but requires additional couplers in order to try to force all of the qubits in a chain to take the same state. This is commonly done by connecting the qubits with a strong ferromagnetic coupling. However, in a sample obtained from the quantum annealer, it is still possible that not all qubits in a chain would have the same state. In that case, a ``decoding'' technique must be utilized to decide the value of the  logical variable corresponding to those qubits. There are various decoding techniques, such as majority vote, and local and global energy minimization \cite{vinci2015quantum}. In this study, we employed local energy minimization for decoding. More specifically, we assigned an effective field to each broken chain, selected the chain with the strongest effective field, set its state to be opposite to that of the direction of the field (to minimize the energy), updated the effective fields, and repeated the process until no broken chains remained. This is a quick method of local error correction. We remark that finding an embedding for a graph is formally known as graph-minor embedding, an NP-hard problem \cite{choi2008minor,choi2011minor} commonly solved by heuristic methods, although for some classes of problems deterministic embeddings can be found \cite{zaribafiyan2017systematic}. 

\begin{figure}[!htbp]  
\captionsetup{width=.9\columnwidth}
\centering
\includegraphics[width=1.0\linewidth]{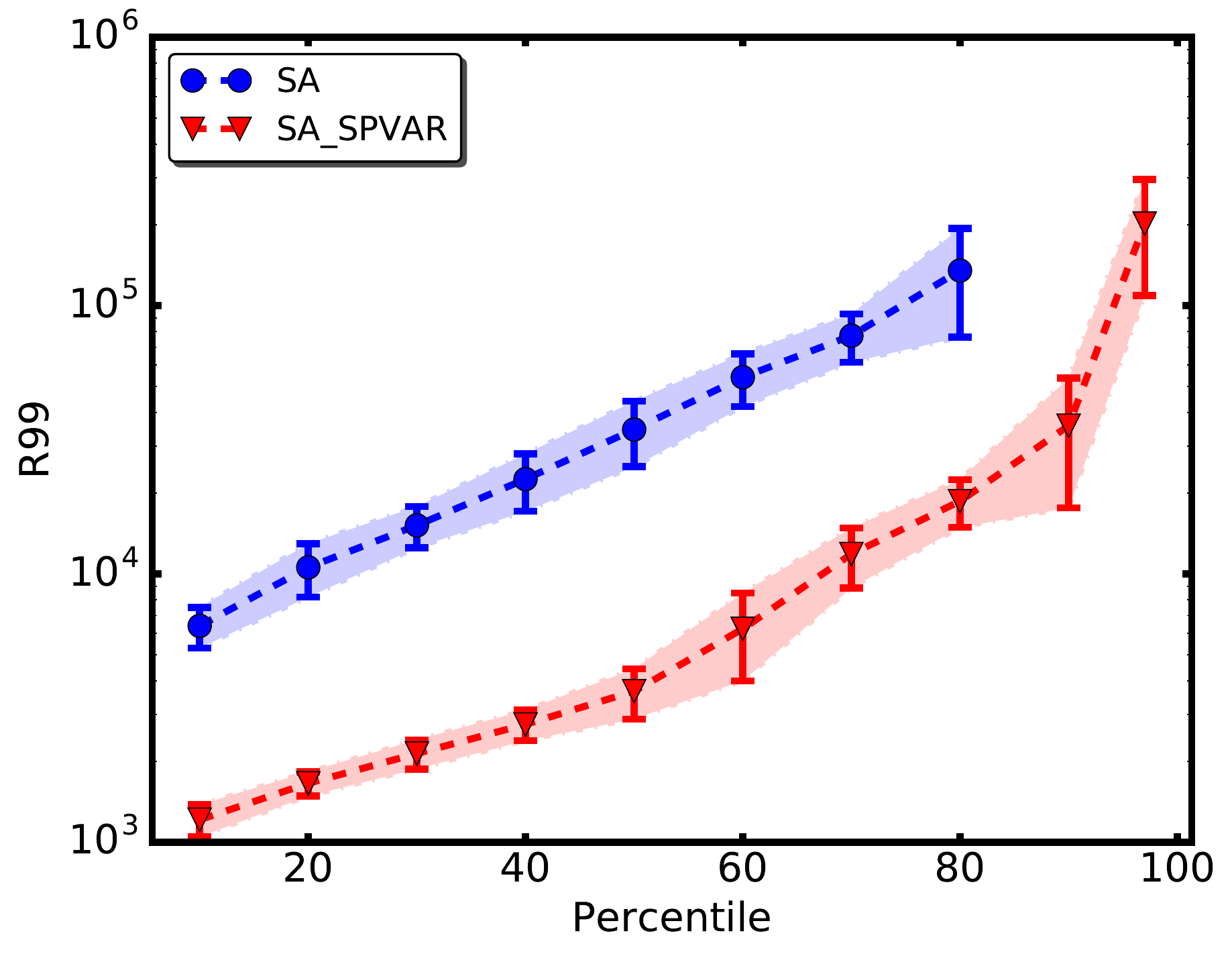} 
\caption{R99 as a function of percentile for the (4,4) fault diagnosis problems, for simulated annealing, with and without SPVAR. The error bars were calculated using bootstrapping.}
\label{fig:fd_vs_percentile}
\end{figure}

We present the success metrics for the fault diagnosis problems in Table~\ref{table:fault_diagnosis}, and the dependence of the R99 on the percentile for SA with and without our method in Figure~\ref{fig:fd_vs_percentile}. 

\begin{table}[!htbp]  
\begin{tabular}{@{\extracolsep{\fill}} lrrrr}
\hline
\hline
\multirow{2}{*}{Sampler} &  \multicolumn{2}{c}{Without SPVAR} & \multicolumn{2}{c}{With SPVAR} \\
                            & Solved & Residual & Solved & Residual \\
\hline
SA (2000 sweeps)            &     21        & 0.167  &  80     & 0.032 \\
SA (20000 sweeps)           &     67       & 0.056  &  96     & 0.005 \\
\hline
SQA (2000 sweeps)           &      0      & 2.055  &   4 & 0.447  \\
SQA (20000 sweeps)            &     2      & 0.348   &   45 & 0.115  \\
\hline
DW                          &     0         & 4.660  &  0   &  1.278  \\
DW + PP                     &     19   & 0.189  &  88 & 0.017  \\
\hline
\hline
\end{tabular}
\caption{Success metrics for the (4,4) fault diagnosis problems for three samplers: simulated annealing (`SA'), simulated quantum annealing (`SQA'), and D-Wave's quantum annealer (`DW'). For each sampler, we present the success metrics for solving using the sampler alone compared with solving using SPVAR, with the same computational effort. For DW, we also show the results with post-processing (`PP'). The success metrics are: `Solved'---the percentage of problems solved; and `Residual'---the mean difference (in percent) in energy between the sample's best result and the best known solution. }
\label{table:fault_diagnosis}
\end{table}

For the fault diagnosis problems, the application of our method results in a large improvement to the success metrics for SA and SQA, as can be seen in Table~\ref{table:fault_diagnosis}. The application of the method significantly improves the quality of solutions found by the quantum annealer, as seen in the residual, but it was still unable to solve any of those problems. We attribute this to the low quality of the samples obtained from the quantum annealer, a result of these problems being extremely hard for it to solve. With the addition of post-processing, the quantum annealer achieves results that are in line with the raw results obtained by SA and SQA, and shows a large improvement when our method is applied. We were not able to apply the same post-processing to SA and SQA, since the post-processing is internal to the quantum annealer's SAPI. 

\subsection{Max-\textbf{\emph{k}}-SAT problems}
\label{sec:results_maxksat}

It is known that the size of the backbone, defined as the number of variables that are true in all of the optima, is an order parameter for Max-$k$-SAT problems \cite{monasson1999determining,achlioptas2001phase}. $k$-SAT and Max-$k$-SAT problems undergo a phase transition (second-order), with the order parameter defined as the size of the backbone, at some critical point. At this critical point,  \mbox{Max-$k$-SAT} problems undergo an ``easy-hard'' phase transition. It is natural to ask how our method would perform near this phase transition for Max-$k$-SAT problems. The critical point has been proven to be at $\phi_c=1$ for $k=2$ \cite{chvatal1992mick, de1992random, goerdt1996threshold}, and has been shown to be at $\phi_c\simeq 4.267$ for $k=3$, where $\phi$ is the ratio of the clauses to literals \cite{mezard2002random,mezard2002analytic,mertens2006threshold}.

\begin{figure*}[!htbp]  
    \centering
    \subfloat[Max-2-SAT]{
        \includegraphics[width=0.45\textwidth]{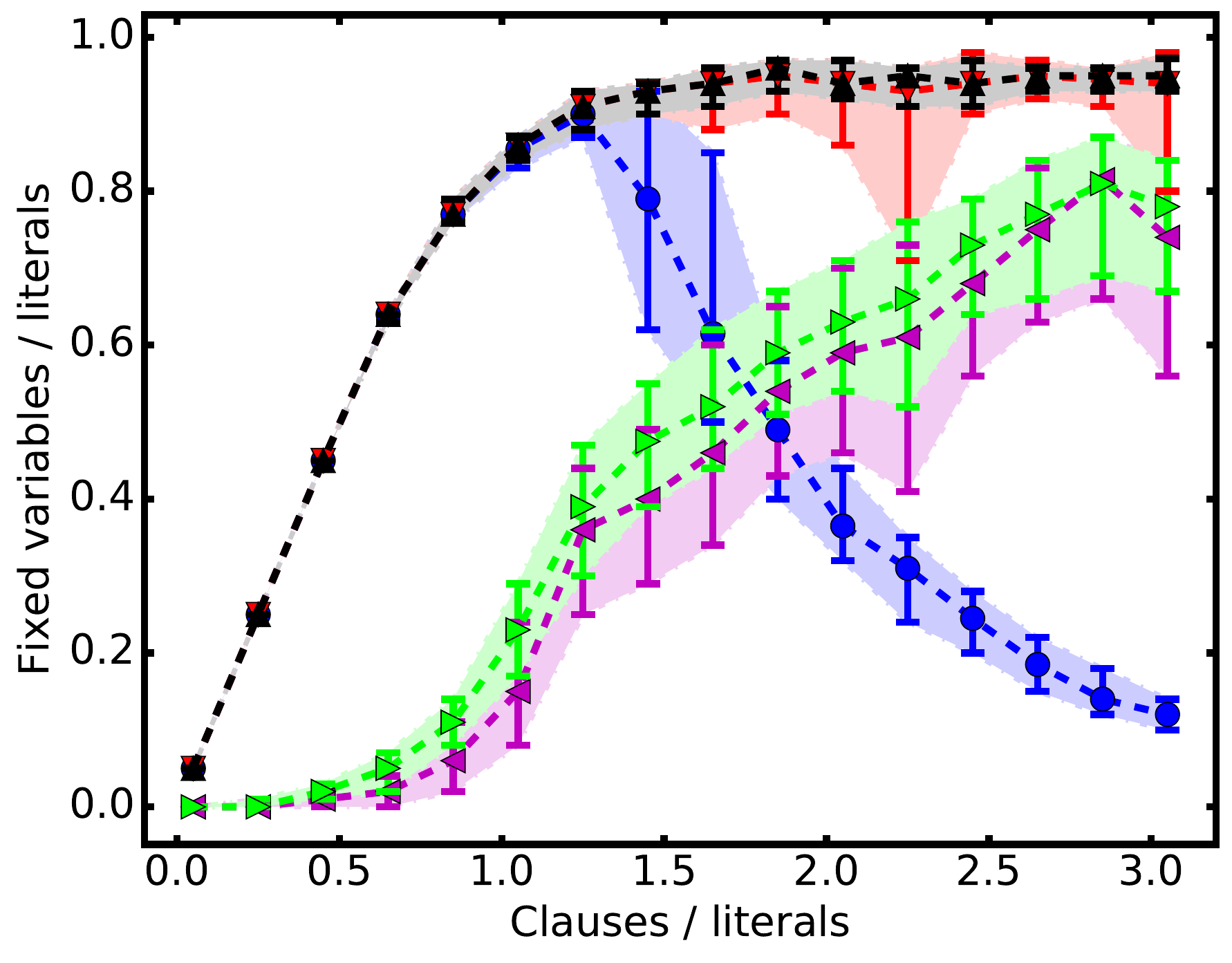} 
	\label{fig:max2sat_num_fixed_qubits} 
    }
    \quad
    \subfloat[Max-3-SAT]{
         \includegraphics[width=0.45\textwidth]{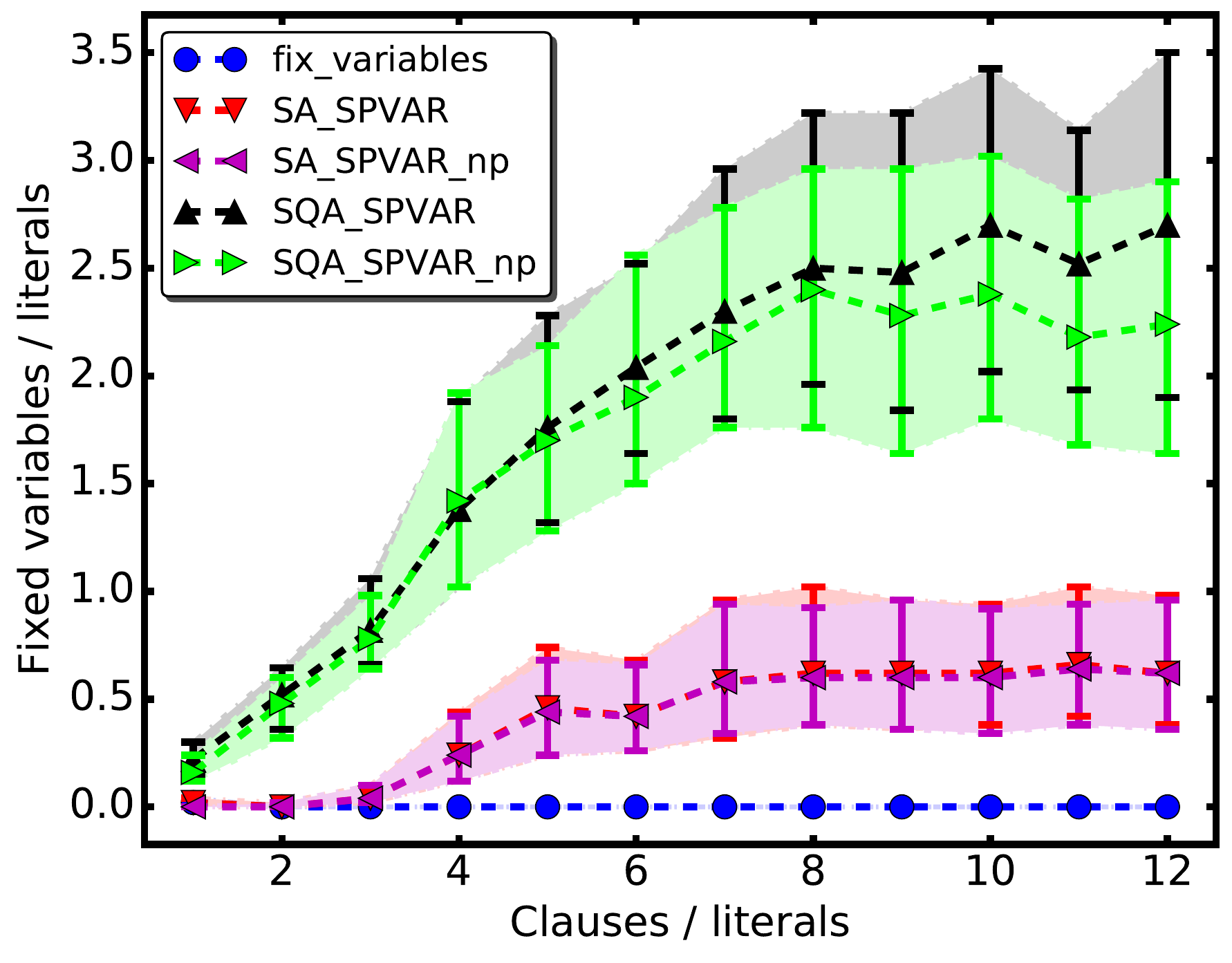} 
         \label{fig:max3sat50_num_fixed_qubits}  
	}
\caption{Number of fixed variables for Max-$k$-SAT problems. {\bf(a)} The median number of fixed variables for Max-2-SAT problems with 100 literals as a function of the number of clauses. `SA\textunderscore SPVAR'  refers to simulated annealing run in conjunction with SPVAR, `SQA\textunderscore SPVAR' refers to simulated quantum annealing run in conjunction with SPVAR, and `fix\textunderscore variables' refers to persistence-based fixing of variables, via the $fix\textunderscore variables$ function in D-Wave's SAPI 2. The suffix `np' refers to switching off the optional calls to $fix\textunderscore variables$ in Multi-start SPVAR. {\bf(b)} The median number of fixed variables for Max-3-SAT problems with 50 literals as a function of the number of clauses. In both figures, the means were taken over 50 random instances with the respective number of clauses. }
\label{fig:maxksat_num_fixed_qubits}
\end{figure*}

\begin{figure*}[!htbp]  
    \centering
   \subfloat[]{
         \includegraphics[width=0.45\textwidth]{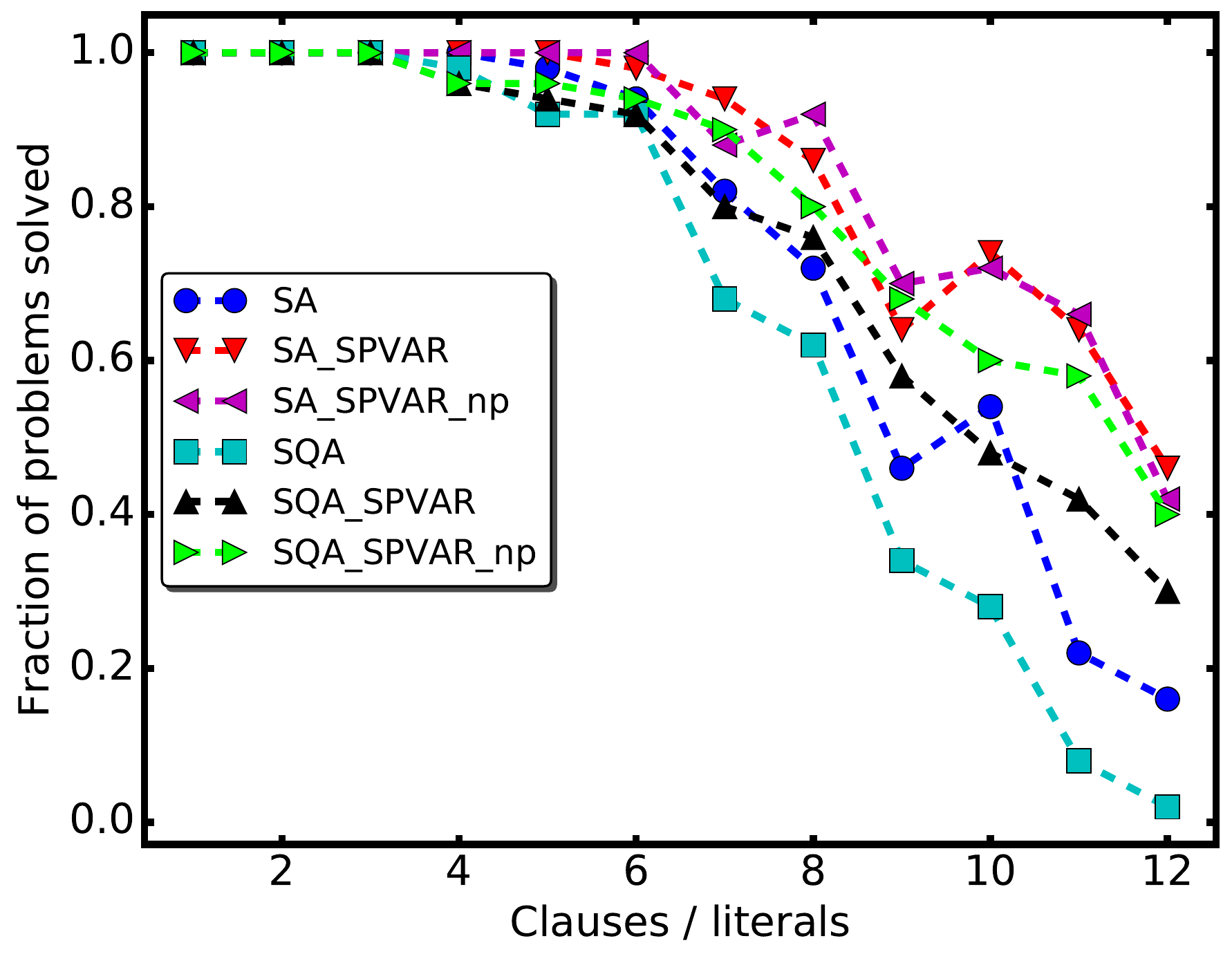} 
         \label{fig:max3sat50_fraction_problems_solved}  
	}
    \quad
	\subfloat[]{
        \includegraphics[width=0.45\textwidth]{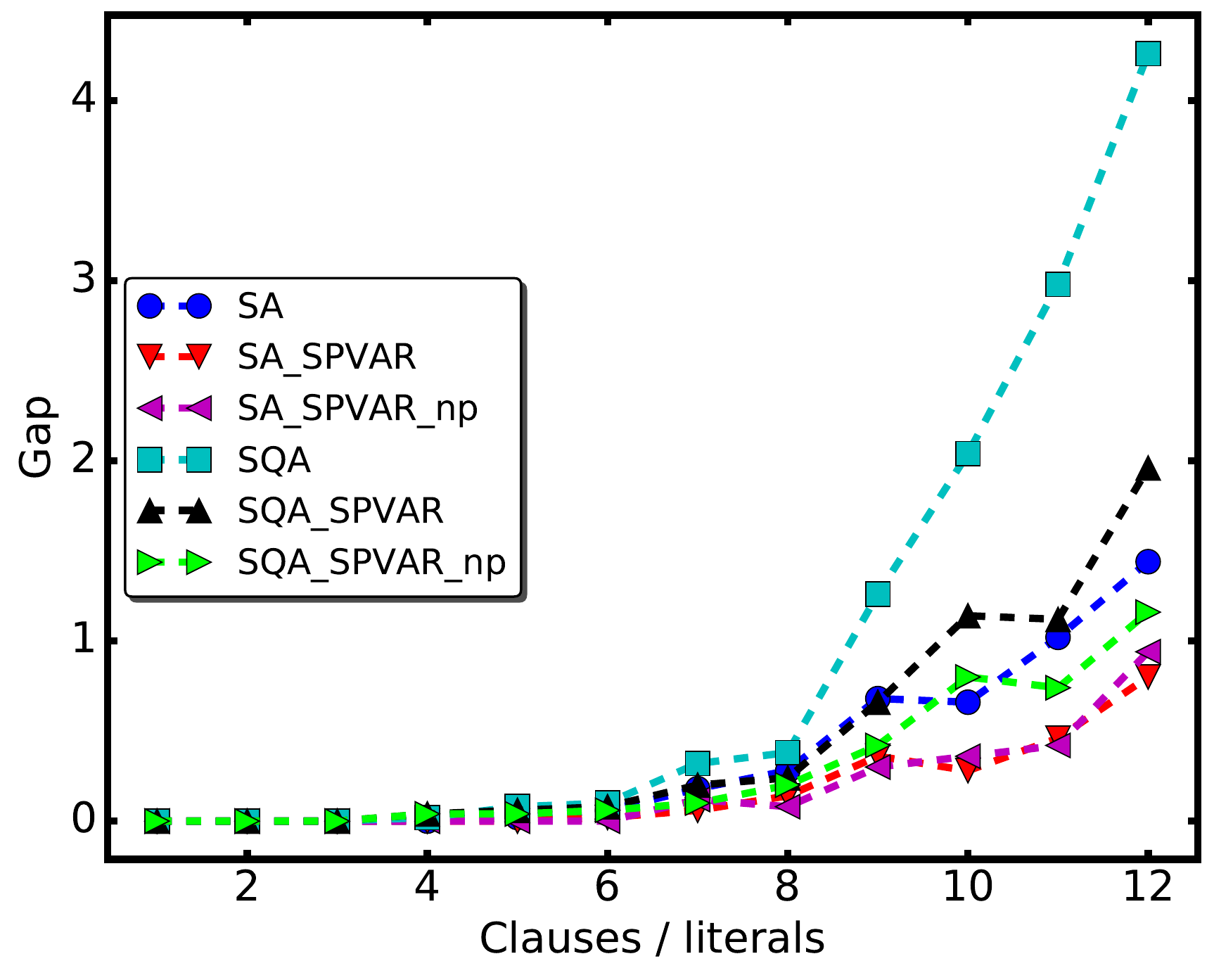} 
	\label{fig:max3sat50_gap} 
    }
\caption{Success metrics for Max-3-SAT problems with 50 literals. {\bf(a)} The fraction of problems solved, as a function of the number of clauses. {\bf(b)} The median gap, which is the energy difference between the best solution found and the best known solution, as a function of the number of clauses. For clarity, error bars are not shown. In both figures, each point represents data from 50 random instances with the respective number of clauses. }
\label{fig:max3sat}
\end{figure*}

We generated random sets of Max-$k$-SAT problems with a given number of literals and a varying number of variables. For the Max-3-SAT instances, the resulting objective function is of order three, necessitating the addition of auxiliary variables and corresponding terms in order to reduce it to a quadratic objective function, since the samplers we used were all implemented to solve quadratic unconstrained problems. A comparison of the number of fixed variables for different methods is presented in 
Figure~\ref{fig:maxksat_num_fixed_qubits}, for Max-2-SAT and Max-3-SAT problems. In addition, for the Max-3-SAT problems, the fraction of problems solved and the gap are presented in Figure~\ref{fig:max3sat}. 


Based on our observations, our method performs well near the phase transition, and continues to perform well as the number of clauses over literals is increased, which corresponds to an increase in hardness. This can be seen both in the number of fixed variables remaining high to the right of the phase transition in Figure~\ref{fig:maxksat_num_fixed_qubits}, as well as in the fact that the method continues to provide a boost to success metrics, as seen in Figure~\ref{fig:max3sat}.

To the left of the phase transition, it is easy to satisfy all of the clauses, and the problems are expected to have a large number of ground states, making them easy to solve for local search methods, such as tabu 1-opt search. In this regime, for $k=2$, the method without the optional calls to $fix\textunderscore variables$ fixes very few variables. This could be explained by the intuition that, due to the large number of ground states, low-energy samplers tend to give a large number of distinct states with low overlap between them. However, this is the regime in which $fix\textunderscore variables$ performs well. To the right of the phase transition, $fix\textunderscore variables$ rapidly breaks down, whereas our method continues to fix a large number of variables. It appears that using our method in conjunction with $fix\textunderscore variables$ allows one to benefit from the strengths of each.  

The number of variables that are fixed by our algorithm is related to the size of the backbone, which we refer to as an approximate backbone. For this reason, it is not surprising that we observed a distinct change in behaviour of the number of fixed variables near the phase transition. 

For $k=3$, $fix\textunderscore variables$ is not able to fix any variables. We hypothesize that the reduction of the third-order polynomial to a quadratic one results in a structure that is not well suited to the type of persistence fixing that $fix\textunderscore variables$ employs. However, it is still able to provide a small increase in the number of variables fixed when used in conjunction with our method. For $k=2$, SA and SQA perform similarly, but for $k=3$, SQA fixes substantially more variables than SA. There are at least two factors that might contribute to this. Firstly, on the left of the phase transition, the curve for SQA is much steeper for $k=3$ than for $k=2$. This might seem unintuitive, due to the observation that in this regime samples are expected to contain states with low overlap. However, recent work on SQA \cite{matsuda2009ground} and the experimental results on the D-Wave quantum annealer \cite{mandra2017exponentially} suggest that their samples are biased towards few of the ground states, which might help to explain this. Secondly, it has been argued that higher-order polynomials provide a rougher landscape, which could lead to an advantage for SQA over SA, at least if the barriers are thin \cite{denchev2015computational}.

\section{Conclusions and future work}
\label{sec:conclusions}

We set out two objectives for this work: to study the performance of SPVAR on hard problems with different samplers, and to study what classes of problems or samplers might not benefit from the application of our method. The results show that for almost all problems studied, even extremely hard problems, the application of our method results in a significant improvement in the success metrics, for all samplers studied. 

In general, SPVAR can result in a reduction in the number of variables. In some cases, the reduction of variables might also result in a computationally simpler problem (e.g., a simplified topology). Therefore, when comparing the numerical effort as a function of the size of the input of a particular heuristic to the same heuristic with SPVAR, an advantage in the scaling can occur via an improved exponential (due to the simplification of a problem) or a reduced prefactor (due to the reduction of the number of variables). Figure \ref{fig:chimera_vs_size} depicts a case where the scaling of the algorithm is improved via the addition of SPVAR. A characterization of the conditions and mechanism under which SPVAR can result in an improvement in scaling would be an interesting future study, however, larger problems would be needed to carefully probe the asymptotic regime. 

We have also identified several types of problems that appear to challenge SPVAR. Firstly, zero-bias problems tend to have large correlation lengths, such that clusters of spins cannot be fixed locally. The reduced-degeneracy Chimera graph problems with zero bias fall under this category. As shown in \mbox{Section~\ref{sec:results_chimera}}, setting the \textit{elite\textunderscore threshold} adaptively can help with this. It is also possible to iteratively apply SPVAR sequentially, which leads to the problems gradually accumulating increasing numbers of variables with non-zero bias, which makes it easier to fix variables, as shown in \cite{karimi2017boosting}. 

Secondly, problems in which the ``cost'' in value due to an incorrectly fixed variable is large compared with the range of couplers. Increasing \textit{num\textunderscore starts} can help with this, by reducing the risk of fixing variables incorrectly. Finally, if a sampler gives a low-quality result, it is possible that the application of our method will not improve the results, although it is not expected to make them worse. By setting \textit{elite\textunderscore threshold} adaptively, it appears to be possible to improve the results, although the risk of fixing variables incorrectly in this case could be high. 

In the future, it might be worthwhile to study the performance of a multi-start iterative application of SPVAR, as well as the possible application of a similar algorithm to other discrete optimization problems, such as mixed-integer programming. 

\section{Acknowledgements}

The authors would like to thank Marko Bucyk for editing the manuscript, Zheng Zhu for the use of his implementation of the \emph{borealis} algorithm (PTICM) \cite{zhu2016borealis}, Matthias Troyer's group for the use of their implementation of the simulated annealing (SA) algorithm \cite{isakov2015optimised}, and Alex Selby for the use of his implementation of the Hamze--de Freitas--Selby (HFS) algorithm, available for public use on GitHub \cite{selby2013github}. We would also like to thank Andr\'{e} Abram\'{e} and Djamal Habet for the use of their Max-$k$-SAT solver \emph{ahmaxsat} \cite{abrame2015ahmaxsat} and Chuan Luo for the use of the Max-$k$-SAT solver CCLS \cite{luo2015ccls}. In addition, we would like to thank Wenlong Wang for supplying us with the 3D spin glass instances, Sergio Boixo for supplying us with the weak-strong cluster problem instances, and Alejandro Perdomo-Ortiz for providing us with the fault diagnosis problem instances. This work was supported by 1QBit and Mitacs. 

The research of HGK was supported by the National Science Foundation (Grant No.~DMR-1151387) and is based upon work supported in part by the Office of the Director of National Intelligence (ODNI), Intelligence Advanced Research Projects Activity (IARPA), via MIT Lincoln Laboratory Air Force Contract No.~FA8721-05-C-0002. The views and conclusions contained herein are those of the authors and should not be interpreted as necessarily representing the official policies or endorsements, either expressed or implied, of ODNI, IARPA, or the U.S.~Government.  The U.S.~Government is authorized to reproduce and distribute reprints for Governmental purpose notwithstanding any copyright annotation thereon. HGK would like to thank Alibi B.~Room for inspiration.


\begin{appendices}

\section{Parameters used}
\label{appendix:parameters}

In Table~\ref{table:parameters}, we list the parameter values used for the benchmarking, for each problem set. In addition, \textit{fixing\textunderscore threshold} was always set to 1.0, as was \textit{correlation\textunderscore threshold}, and the chain repairing method used for quantum annealer runs was local energy minimization. The thresholds column shows first the \textit{elite\textunderscore threshold} and then the  \mbox{\textit{correlation\textunderscore elite\textunderscore threshold}}.

For the non-zero-bias problems, we used half of the \mbox{\textit{total\textunderscore sample\textunderscore size}} for the SPVAR fixing step, and half of the sample size to solve the problems. For the zero-bias problems, we used $40\%$ of the total sample size for the correlation-based pre-fixing step, and divided the rest equally between the fixing step and solving step. 

\begin{table}[!htbp]  
\centering
{\scriptsize \begin{tabular*}{\columnwidth}{@{\extracolsep{\fill}} lrrrr}
\hline
\hline
	Problems &  \textit{elite\textunderscore threshold}s & \textit{num\textunderscore starts} & \textit{total\textunderscore sample\textunderscore size} & \textit{num\textunderscore sweeps} \\
\hline
	Weak-strong & 0.2, 0.2 & 20 & 100--500 & 20 $\times 10^3$ \\
	Red. deg, Chimera & 0.2, 0.2 & 20 & 1000 & 20 $\times 10^3$ \\
	3D spin glass & 0.2, 0.3 & 20 & 500 & 20 $\times 10^3$ \\
	3D spin glass (PTICM) & 0.2, \hspace{0.25em}---\quad & 10 & 500 & 10 $\times 10^3$ \\
	Fault diagnosis & 0.1, 0.4 & 40 & 500 & 2, 20 $\times 10^3$ \\
	Max-$k$-SAT & 0.2, 0.2 & 40 & 500 & 20 $\times 10^3$ \\
\hline
\hline
\end{tabular*}}
\caption{Parameter values of Multi-start SPVAR and the various problem sets used in the benchmarking}
\label{table:parameters}
\end{table}
%


\end{appendices}

\clearpage

\bibliographystyle{apsrevtitle}

\bibliography{Effective_Optimization_Sample_Persistence}

\end{document}